\documentclass[journal]{IEEEtran}
\usepackage{amsmath,amssymb,amsthm,epsfig,color,empheq,graphicx,graphics,balance}
\usepackage{enumerate,url,wasysym,epstopdf,enumitem,array}
\usepackage{xcolor}
\usepackage{amsfonts}
\usepackage{adjustbox}
\usepackage{multirow}
\usepackage{cite}
\usepackage{accents}

\usepackage{algorithm2e} 
\usepackage{float} 
\usepackage{booktabs}   

\newtheorem{lemma}{Lemma}
\newtheorem{corollary}{Corollary}

\newtheorem{remark}{Remark}
\newtheorem{example}{Example}

\newcommand \expect{\mathbb{E}}

\newcommand \bone{\mathbf{1}}
\newcommand \ba{\mathbf{a}}
\newcommand \bb{\mathbf{b}}

\newcommand \bt{\mathbf{t}}

\newcommand \mcF{\mathcal{F}}

\newcommand \mcN{\mathcal{N}}

\usepackage{booktabs}
\usepackage{graphicx}

\begin{document}

\title{Frequency-Domain Characterization of\\
Load Demand from Electrified Highways}

\author{Ashutossh Gupta,~\IEEEmembership{Graduate Student Member,~IEEE}, 
        Vassilis Kekatos,~\IEEEmembership{Senior Member,~IEEE},
        Ruoyu Yang,~\IEEEmembership{Student Member,~IEEE},
        Dionysios Aliprantis,~\IEEEmembership{Fellow,~IEEE}, and
        Steven Pekarek,~\IEEEmembership{Fellow,~IEEE}
\thanks{Manuscript received September 2, 2025; revised February 13 and April 23, 2026; accepted June 6, 2026. This work was partly supported by the U.S. National Science Foundation under grants 2500682 and 1941524. A.~Gupta, V.~Kekatos, R.~Yang, D.~Aliprantis, and S.~Pekarek are with the Elmore Family School of ECE, Purdue University, West Lafayette, IN 47906, USA. Emails:\{gupta799,kekatos,yang2729,dionysios,spekarek\}@purdue.edu.}
\thanks{Color versions of one or more of the figures in this paper are available online at {http://ieeexplore.ieee.org}.}
\thanks{Digital Object Identifier TPWRS-01549-2025}
}	
	
\markboth{IEEE TRANSACTIONS ON POWER SYSTEMS (accepted June 6, 2026)}{IEEE TRANSACTIONS ON POWER SYSTEMS (accepted June 6, 2026)}

\maketitle

\begin{abstract}
Electrified roadways (ER) equipped with dynamic wireless power transfer (DWPT) capabilities can patently extend the driving range and reduce the battery size of electric vehicles (EVs). However, due to the spatial arrangement of the transmitter coils in the ER, the DWPT load exhibits frequency content that could excite power system frequency dynamics. In this context, this work aims to study the spectrum of DWPT loads under different traffic conditions. Under simplifying assumptions, we develop statistical models to identify the location and relative magnitude of DWPT load harmonics. Our analysis reveals that the fundamental frequency depends on ER coil spacing and average EV speed. In the worst-case yet unlikely scenario that EVs move in a synchronized fashion, the amplitude of harmonics scales with the EV count. On the contrary, when EVs move freely, harmonics scale with the square root of the EV count. Platoon formations can accentuate harmonics. The spectral content around harmonics decreases in magnitude and increases in bandwidth with the harmonic index. The load of a single EV moving at a time-varying speed can be modeled as a frequency-modulated (FM) signal. Despite the simplifying assumptions, the derived models offer valuable insights for ER planners and grid operators. Dynamic simulations of a modified WECC model with DWPT loads synthesized from realistic EV trajectories and ER specifications corroborate some of these insights.
\end{abstract}


\section{Introduction} \label{sec:intro}
\allowdisplaybreaks
Electrified roadways (ERs) with DWPT capabilities have been proposed to power EVs during transit. The power can be used directly for vehicle propulsion and to charge a battery if needed. DWPT-enabled ERs are expected to offer multiple benefits, such as extended EV range, reduced battery size, and reduced overall battery charging time~\cite{Diala19}. DWPT systems may also contribute to increasing EV adoption along with potential cost benefits~\cite{Diala22}. {Currently, the DWPT technology has been proposed for vehicles moving on fixed routes, such as public transit buses and regional-haul Class 8 trucks, as it simplifies infrastructure planning and utilization~\cite{borlaung2021heavy-duty} \cite{sutton2024effectiveness}.}

From a power grid perspective, with growing EV penetration, stationary EV charging is expected to increase the load on distribution systems significantly~\cite{zhu2020grid}. For instance, reference~\cite{el2022impact} shows that even if 11\% of the total number of heavy-duty EVs in Texas charge simultaneously, they would create significant voltage violations. While stationary EV charging loads are concentrated at specific grid buses, DWPT-enabled ERs can distribute EV load across multiple buses.


The DWPT system relies on an array of transmitters (Tx) embedded in the ER that supply power to receivers (Rx) installed on EVs. A segment of an ER draws power from the power grid through a substation. Depending on the ER segment length and anticipated DWPT traffic, the substation can be at the transmission or distribution voltage levels. The substation may serve a segment of an electrified highway and/or additional loads. In most DWPT implementations, successive Tx coils on the ER are separated by a gap~\cite{vatan2025receiver},~\cite{vatan2026roadway},~\cite{Cai_longtrack_2022}. Owing to the spatial arrangement of the Tx coils, the power consumed by each EV exhibits a time-varying, near-periodic waveform. The total DWPT load consumed at the substation equals the sum of the EV loads plus losses. This work aims to model the total DWPT load consumed at this substation in the frequency domain. 

Previous work predicts the average power consumption of the ER infrastructure~\cite{sauter2024power}. Such DWPT load predictions are crucial for system operators to conduct unit commitment, economic dispatch, and power flow studies, as well as determine market prices~\cite{xie2016sequential}. Using similar DWPT load predictions, researchers have also designed battery storage systems and control mechanisms to regulate voltage on the distribution grid feeding the ER~\cite{newbolt2023diverse}. Reference \cite{liu2023bilevel} studies the optimal location of DWPT lanes, and \cite{newbolt2024feasibility} investigates the vehicle-to-grid power transfer capabilities during blackouts. 

While hourly-averaged DWPT load profiles are essential for grid scheduling, they cannot predict the impact of ERs on power grid frequency dynamics. Because stationary EV loads are approximately constant over finer time scales, they have not been a concern for exciting power system oscillations. On the contrary, dynamic EV loads are time-varying and thus pose the risk of introducing oscillations into the power grid. To model the dynamic impact of a DWPT system, its load demand must be studied at the second timescale. By assessing the shape of the DWPT frequency spectrum, grid operators can evaluate whether ERs could introduce frequency oscillations into transmission systems. 

Transmission system operators are interested in identifying possible sources of sustained low-frequency oscillations, also known as \textit{inter-area oscillations}~\cite{gupta2025incorporating}. These oscillations generally occur at frequencies below 2~Hz and can lead to widespread blackouts if not properly damped; see the recent 2025 blackout in the Iberian Peninsula~\cite {batlle2025hopefully}. Such oscillations can be triggered due to malfunctions in switching of power electronic devices~\cite{Cheng2023subsynchronous}, \cite{Alizadeh2025Ontario}; sudden loss of load and generators~\cite{lian2017universal}; as well as pulsating electrical loads~\cite{mishra2025understanding}. Identifying the sources of these oscillations based on real-time measurements is time-intensive. Therefore, it becomes pertinent for power system operators to pre-identify potential sources of such oscillations. This work aims to develop models that determine the frequency spectrum of the total power consumed by a DWPT system under various traffic conditions. Such models are helpful for grid operators and ER designers to understand the impact of DWPT systems on power system dynamics and develop appropriate solutions. 

{Studies investigating the effects of DWPT systems on the power grid utilize hourly traffic flow data to estimate minute-scale power demand profiles and study voltage violations across a distribution feeder~\cite{Travis23}. At this timescale, the effect of Tx coil segmentation can be safely ignored. Reference~\cite{Diala19} incorporated the effect of coil segmentation in determining DWPT load demand under different EV penetration levels. Nonetheless, the DWPT spectrum was computed by averaging across diverse traffic conditions, with EV spacings drawn from a distribution inferred from year-long traffic flow data. Recent work~\cite{newbolt2024load} proposes a discrete-time convolution-based method to simulate DWPT loads at a finer timescale, but does not characterize their frequency-domain characteristics.}


This work contributes to the ensuing fronts:
\begin{itemize}
\item[\emph{c1)}] We develop statistical models for the DWPT load consumed at a substation serving an ER segment (Sec.~\ref{sec:modeling}). The DWPT load is modeled as a continuous-time signal and analyzed in the time and frequency domains. Based on these models, we identify the location and amplitude of the DWPT load's frequency harmonics and study its \emph{total harmonic content} (THC), defined as the ratio of the non-DC to the DC components. Our analysis establishes scaling laws to inform ER designers and grid operators about the potential impact of DWPT loads on grid frequency dynamics.
\item[\emph{c2)}] Our findings on the DWPT spectrum are summarized as follows (Secs.~\ref{sec:cons&equal_vel}--\ref{sec:cons&uneq_vel}):
\begin{itemize}
\item Depending on the ER coil spacing and EV speed, the DWPT load can exhibit frequency content below $2$~Hz, that is, within the range of inter-area oscillations.
\item Assuming EVs move at a common and constant speed, the highest THC occurs when EVs move in a synchronized fashion. In this worst-case yet unlikely scenario, both the average (DC) DWPT load and its harmonics scale with the number of EVs $N$. (Scenario \emph{S$_1$})
\item When EVs move asynchronously, harmonics scale with $\sqrt{N}$. (Scenario \emph{S$_2$})
\item When EVs form platoons with $Q$ EVs per platoon, harmonics scale with $\sqrt{QN}$. (Scenario \emph{S$_3$})
\item If EVs move at constant yet unequal speeds, the DWPT load becomes non-periodic. Its spectrum consists of lobes centered around harmonics associated with the average speed. The lobes decrease in magnitude and increase in bandwidth for higher harmonics. {Harmonics scale with $\sqrt{N}$ as in \emph{S$_2$}.} (Scenario \emph{S$_4$})
\item {If EVs move at time-varying speeds, their individual DWPT loads can be modeled as \emph{frequency-modulated} (FM) signals. Spectra consist of main lobes at harmonics corresponding to the average speed. Each main lobe is accompanied by multiple sidelobes arising due to deviations of speed from the average. (Scenario \emph{S$_5$})}
\end{itemize}
\item[\emph{c3)}] Our findings are validated numerically as follows:
\begin{itemize}
    \item Periodograms of randomly sampled DWPT loads resemble the analytically derived spectra (Sec.~\ref{sec:matlab_tests}). 
    
    \item {DWPT loads generated by the open-source traffic-flow simulator SUMO~\cite{SUMO18} demonstrate that the analyses under \emph{S$_1$} and \emph{S$_4$} explain the periodograms obtained under free-flow and congested traffic, respectively (Sec.~\ref{sec:sumo}).}
    
    \item {DWPT spectrograms (periodograms plotted across time) demonstrate that free-moving traffic yields mild THC, whereas slower-speed heavier traffic could lead to synchronization reaching THC of 18\%; see Sec.~\ref{sec:sumo}.}
    
    \item {Feeding DWPT loads obtained from SUMO to the modified WECC 179-bus system demonstrates that during congestion, the DWPT load could excite inter-area oscillations (Sec.~\ref{sec:dynamics_simu}). DWPT loads due to free-flowing traffic are unlikely to induce frequency oscillations. Spreading loads across a power system can exacerbate oscillations.}
\end{itemize}
\end{itemize}

\begin{table}[t]
\centering

\caption{Nomenclature} \label{tab:nomenclature}
{\fontsize{8pt}{11pt}\selectfont
\begin{tabular}{|l|l|}
\hline
\hline
  \textbf{Symbol}   &  \textbf{Meaning} \\
     \hline
     \hline
     $L$ & length of ER segment \\
     \hline
     $N$ & number of EVs \\
     \hline
     $\ell_T,\ell_R$ & length of Tx/RX coils \\
     \hline
     $d$ & spacing between successive Tx coils \\
     \hline
     $D = \ell_T + d$ & coil segment length \\
     \hline
     $K = L/D$ & number of coil segments \\
     \hline
     $a_n$ & spatial power density absorbed by EV$_n$   \\
     \hline
     $x_n$ & position of head of Rx of EV$_n$ \\
     \hline
     $O(x_n)$ & Rx/Tx overlap for EV$_n$ at position $x_n$ \\
     \hline
     $\bar{p}_n(x_n)$ & power consumed by EV$_n$ at position $x_n$ \\
     \hline
     $v_n(t)$ & speed of EV$_n$ \\
     \hline
      $p_n(t)$ & power consumed by EV$_n$ at time $t$ \\
     \hline
      $\bar{a}$ & rated power density of an ER  \\
     \hline
      $p(t)$ & total power consumed by a DWPT system  \\
     \hline
      $t_n$ & timing of EV$_n$  \\
     \hline
      $h_T(t)$ & train of trapezoidal pulses of period $T$  \\
     \hline
      $\omega_0=2\pi/T$ & fundamental frequency of $h_T(t)$ \\
     \hline
      $c_m$ & $m$-th Fourier series coefficient of $h_T(t)$  \\
     \hline
      THC$_h$,THC$_p$  & THC of $h_T(t)$, $p(t)$  \\
     \hline
      $\tilde{c}_m$ & $m$-th FS coeffient of $p(t)$  \\
     \hline
      $P(\omega)$ & Fourier transform of $p(t)$  \\
     \hline
     $\delta(\omega)$ & Dirac delta function  \\
     \hline
      $\mu_p, R_p(\tau)$ & mean and autocorrelation of $p(t)$  \\
     \hline
      $S_p(\omega) = \mcF[R_p(\tau)]$ & Power spectral density of $p(t)$  \\
     \hline
      $\mu_a,\sigma_a^2$ & mean and variance of $a_n$   \\
     \hline
      $Q$ & number of EVs per platoon   \\
     \hline
      $\lambda_a$ & mean vehicle arrival rate   \\
     \hline
      $T_n = D/v_n$ & fundamental period of EV$_n$   \\
     \hline
       $\mu_v,\sigma_v^2$ & mean and variance of speeds $v_n$   \\
     \hline
      $\mu_{\omega},\sigma_{\omega}^2$ & mean and variance of  $\omega_n$   \\
     \hline
     $\hat{p}(t)$ & DWPT load realization of duration $T_p$  \\
     \hline
     $\hat{S}_p(\omega)$ & periodogram of $\hat{p}(t)$  \\
     \hline
     $\bar{v}_n$,$\tilde{v}_n(t)$ &  mean EV speed, deviations from mean speed\\
     \hline
     $\omega_s = 2\pi/D$ & spatial fundamental frequency \\
     \hline
     \hline
     
\end{tabular}
}
\end{table}

\section{Time-Domain Model of DWPT Load Demand}\label{sec:modeling}
This section models the total power consumed by all EVs powered by a DWPT-enabled ER. Hereafter, we will refer to DWPT-enabled EVs simply as EVs. {We assume the DWPT technology is available on a single lane of the ER as suggested by many existing DWPT studies~\cite{bafandkar2025charging}. Moreover, without loss of generality, we consider one direction of the ER, e.g., northbound alone, rather than both northbound and southbound.} We consider an ER segment of length $L$ that is powered by a substation. We are interested in modeling the power demand at that substation in the time and frequency domains.   

\begin{figure}[t]
\centering
\includegraphics[width=0.95\columnwidth]{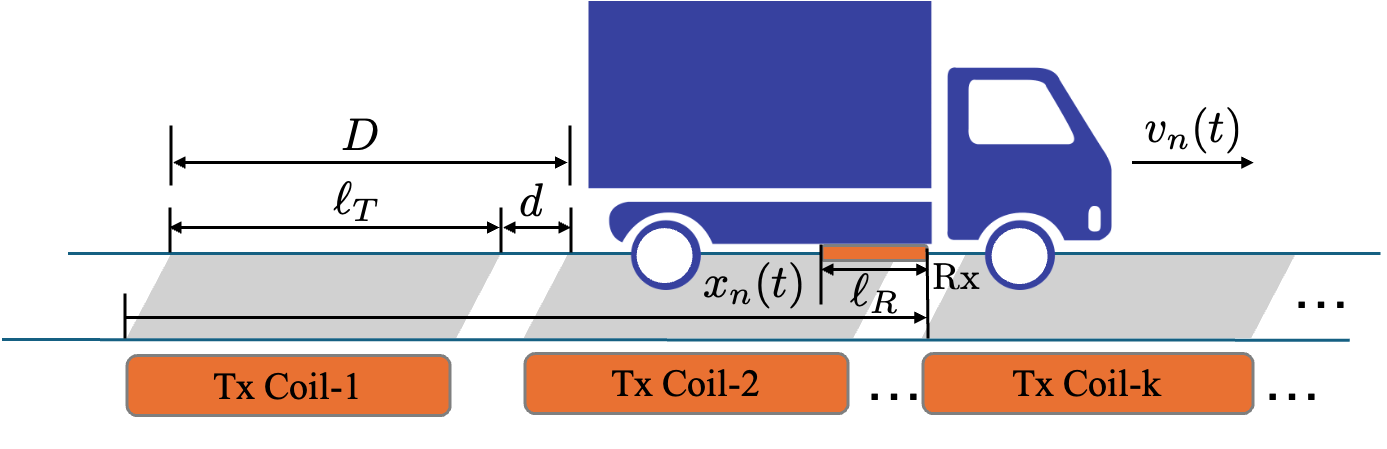}
\caption{In a DWPT-enabled ER, transmitters Tx of length $\ell_T$ and gaps of length $d$ are arranged periodically, every $D = \ell_T + d$~meters. An EV draws power when its Rx of length $\ell_R$ overlaps with the Tx coils.}
\label{fig:dwpt1}
\end{figure}

The ER consists of Tx coils, each of length $\ell_T$, laid in a linear arrangement; see Fig.~\ref{fig:dwpt1}. Consecutive coils are spaced $d$ meters apart due to road construction specifications. Each Tx coil and the subsequent gap $d$ will be termed a \emph{coil segment} of length 
\begin{equation}\label{eq:D}
D=\ell_T+d.     
\end{equation}
The pattern of coil segments is repeated in the direction of traffic flow with period $D$. Coil segments are indexed by $k =0,\dots, K-1$, where $K={L}/{D}$. Electric vehicles are instrumented with Rx of length $\ell_R$. Although our analysis assumes $\ell_R$ to be fixed across EVs, it can be extended to capture EVs of different types, and hence, different Rx lengths. 

Suppose the ER segment serves $N$ EVs, indexed by $n=1,\dots,N$. We first model the power consumed by a single EV. The power consumed by the $n$-th EV, or simply EV$_n$, depends on two factors: \emph{i)} The overlap of its Rx with the Tx coils, measured in meters; and \emph{ii)} The spatial density $a_n$ (in kW/m) of power absorbed by EV$_n$ per meter of Rx/Tx overlap. Leaving $a_n$ aside for now, let us focus on the Rx/Tx overlap as a function of the EV's position. In detail, if $\ell_T\geq \ell_R$, the overlap of the Rx with the first Tx coil can be computed as
\begin{align}\label{eq:Pi(x)}
O(x_n) = 
\begin{cases}
x_n, & 0\leq x_n< \ell_R\\
\ell_R, & \ell_R \leq x_n < \ell_T\\
\ell_R+\ell_T-x_n, & \ell_T\leq x_n < \ell_T+\ell_R\\
0, & \text{otherwise},
\end{cases}
\end{align}
where $x_n$ is the distance between the front end of the Rx for EV$_n$ and the back end of the first Tx coil; see Fig.~\ref{fig:dwpt1}. Because coil segments are $D$ meters long, the overlap of the Rx with the $(k-1)$-th Tx coil is $O(x_n-kD)$ for $k=0,\ldots,K-1$. Then, the power demand of EV$_n$ can be expressed as
\begin{equation}\label{eq:pn(x)}
\bar{p}_n(x_n) = a_n \sum_{k=0}^{K-1} O(x_n - kD).
\end{equation}

It is clear from \eqref{eq:pn(x)} that the load of EV$_n$ is a periodic function of the EV's position. Nevertheless, to analyze the DWPT load in the frequency domain, we need to model the load of EV$_n$ as a function of time rather than position. Let $x_n(t)$ be the position of EV$_n$ at time $t$, and $v_n(t) = \dot{x}_n(t)$ be its speed. The load of EV$_n$ can then be expressed as a function of time as
\begin{equation}\label{eq:pn(t)}
p_n(t)=\bar{p}_n(x_n(t))=a_n \sum_{k=0}^{K-1} O(x_n(t) - kD).
\end{equation}

\begin{figure}[t]
\centering
\includegraphics[width=\columnwidth]{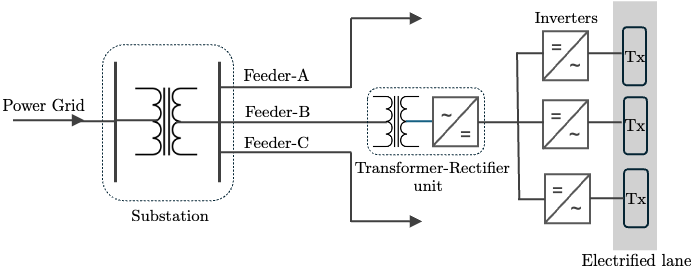}
\caption{The INDOT ER prototype is powered by a substation that steps down grid voltage to $12.47$ kV. Each feeder is connected to a transformer-rectifier unit that converts AC power to DC. DC power is then fed to inverters installed at the side of each Tx coil~\cite{Diala19}. The Tx coils are designed to operate at around $765$ V and can supply $190$ kW of power.}
\label{fig:grid_to_coil}
\end{figure}

\begin{example}\label{ex:indot}
We will often refer to the \textbf{INDOT ER testbed}, a pilot ER project under development by the Indiana Department of Transportation (INDOT). The rightmost lane out of two lanes is instrumented with DWPT technology for a stretch of $400$~m with $\ell_T = 3.66$~m and $d=0.91$~m. A prototype DWPT-enabled EV has also been developed, featuring an Rx of length $\ell_R = 1.8$~m. The speed limit is $v=24.6$~m/s or $55$~mph. This testbed is designed to provide up to $200$~kW to an Rx of length $\ell_R=1.8$~m, yielding a spatial power density of up to $\bar{a}=111.11$~kW/m. {Figure~\ref{fig:grid_to_coil} shows a schematic of the components powering this ER~\cite{PowerOnTheMove}. The distribution substation transformer steps the voltage down to 12.47~kV, which is then supplied to the transformer-rectifier (TR) units via feeders. Each TR unit steps down the voltage level to around 800-900~V as well as converting power from AC to DC to reduce losses. Tx coils are supplied with AC power via dedicated inverters, one per Tx coil. The losses include ohmic losses along the feeder, leakage in the Rx/Tx electromagnetic coupling, and inefficiencies in DC/AC inverters. The INDOT ER pilot has achieved an 85\% Tx-to-Rx power transfer efficiency~\cite{vatan2025receiver}. Note that this is only one possible DWPT architecture, as there are no standardized designs yet.}
\end{example}

The amplitude $a_n$ is determined by the ER power rating and the EV controller. The ER has a rated density $\bar{a}$ of kW/m. This is the maximum power an EV can extract from the ER per meter of Rx/Tx overlap; see blue waveform in Fig.~\ref{fig:trapezoid}. However, depending on the load demand, the EV controller may require less power. This can be implemented in different ways. One option would be to scale down the rated density so that the effective density is set to some $a_n \leq \bar{a}$ in kW/m; see green waveform in Fig.~\ref{fig:trapezoid}. The model in \eqref{eq:pn(t)} captures this multiplicative option. Another option is for the EV controller to saturate its peak density to a value below $\bar{a}$; see red waveform of Fig.~\ref{fig:trapezoid}. In this case, the load of EV$_n$ is a \emph{clipped} rather than a scaled version of the blue waveform. This clipping option simplifies the EV controller implementation and has been implemented on the EV prototype for the INDOT testbed. The model in \eqref{eq:pn(t)} can capture clipping by adjusting the Rx/Tx overlap function per EV. To simplify the analysis, we will henceforth adopt the multiplicative option. 

The total DWPT load demand at the substation serving this ER segment is the sum of EV loads:
\begin{equation}\label{eq:total}
p(t)=\sum_{n=1}^N p_n(t).
\end{equation}

Effects such as Rx/Tx misalignment, voltage drops across the distribution system feeding the ER, and losses could be captured by adding noise terms to \eqref{eq:pn(x)} or \eqref{eq:total}, but are ignored for simplicity. {Inverters and TR units operate at a switching frequency of 85~kHz, which introduces high-frequency content to $p_n(t)$. Because the transmission network is a natural low-pass filter, such content would not excite grid frequency dynamics and can be ignored. Interested readers can refer to~\cite {vatan2025receiver} for precise DWPT voltage and power waveforms.}

We aim to assess the frequency spectrum of $p(t)$. An important summarization of the DWPT load spectrum is the ratio of the undesirable frequency content of $p(t)$ to its DC component. We term this metric \emph{total harmonic content} and define it formally as
\begin{equation}\label{eq:THC_og}
    \text{THC}_p = \frac{\sqrt{\lim_{T \rightarrow \infty} \frac{1}{T} \int_{0}^T \left(p(t)-\Bar{p}\right)^2 dt}}{\Bar{p}} \times 100\%, 
\end{equation}
where $\Bar{p}$ is the time-averaged DWPT load:
\[\Bar{p} = \lim_{T \rightarrow \infty} \frac{1}{T}  \int_{0}^T p(t) dt. \]
The DC component is the actual power consumed by the DWPT load. The non-DC component comprises fluctuations in $p(t)$ about its average value. How are such fluctuations distributed across frequencies? How does THC$_p$ change with traffic conditions? Depending on their frequency content and THC, DWPT loads could excite grid frequency dynamics.

A key challenge in estimating the frequency spectrum of the DWPT load is that EV positions, speeds, and densities are random and time-varying. To this end, we postulate statistical models for $p(t)$ to account for various traffic conditions and derive the corresponding spectra. {We derive the frequency spectrum and THC of the total DWPT load under five models or scenarios:}
\begin{enumerate}
    \item[\emph{S$_1$)}] EVs move at a common constant speed and synchronize in time so their load waveforms add constructively. Although such perfect synchronization is unlikely to occur, it serves as a worst-case scenario.  
    \item[\emph{S$_2$)}] EVs move at a common constant speed and are randomly spaced apart. 
    \item[\emph{S$_3$)}] EVs move at a common constant speed and form platoons. 
    \item[\emph{S$_4$)}] EVs move at different constant speeds and are randomly spaced apart. 
    \item[\emph{S$_5$)}] {Model the spectrum of a single EV moving with time-varying speed.}
\end{enumerate}

\section{EVs Moving at Common Constant Speeds}\label{sec:cons&equal_vel}
In this section, we study the spectrum of the DWPT load under the first three scenarios. These scenarios assume that all EVs move at a common and constant speed, that is $v_n(t)=v$ for all $n$ and $t$. This assumption simplifies the analysis and offers valuable insights. Under constant speed, the position of EV$_n$ at time $t$ can be expressed as
\begin{equation}\label{eq:ConstantSpeedpnt}
x_n(t)=v\cdot (t-t_n)~~\text{for}~~t\geq t_n    \,,
\end{equation}
where $t_n$ is the time EV$_n$ moves over the first Tx coil, and will be termed as the \emph{timing} of EV$_n$. When EV$_n$ moves at a constant speed, its power demand is periodic in time. Precisely, substituting \eqref{eq:ConstantSpeedpnt} into \eqref{eq:pn(t)} yields
\begin{equation}\label{eq:ConstantSpeedpnt2}
p_n(t)= a_n \sum_{k=0}^{K-1}O\left(v\cdot (t-t_n-kT)\right),
\end{equation}
where $T = {D}/{v}$ is the \emph{fundamental period} of $p_n(t)$. It is clear from \eqref{eq:ConstantSpeedpnt2} that each $p_n(t)$ is a scaled and time-shifted version of the periodic signal
\begin{equation}\label{eq:h(t)}
h_T(t)=\sum_{k=-\infty}^{\infty} O\left(v\cdot(t-kT)\right).
\end{equation}

\begin{figure}[t]
\centering
\includegraphics[width=\linewidth]{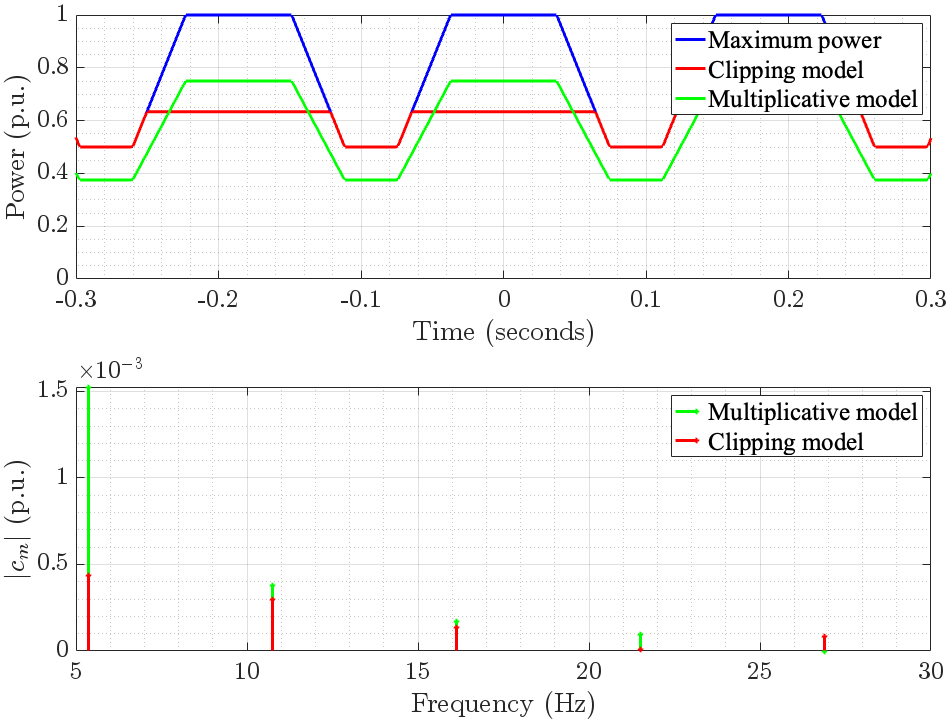}
\caption{\emph{Top:} The DWPT load $p_n(t)$ for an EV with Rx length $\ell_R = 1.8$~m moving at $v = 24.65$~m/s ($55$~mph) on the INDOT ER testbed. The blue and green load waveforms occur when the EV controller follows the multiplicative model with $a_n = \bar{a}$ and $a_n = 0.75 \bar{a}$, respectively. The red load waveform occurs when the EV controller clips the drawn power at $0.63\bar{a}$. The maximum power of the clipped DWPT load is chosen such that the DC component of the red and green waveforms is the same. Additionally, since $\ell_R >d$, the clipped DWPT load is strictly positive at all times. \emph{Bottom:} The FS line spectrum of the red and green load waveform shown on the top panel ($c_0 = 0.6$~p.u.). The spectrum peaks at harmonics $mf_0$ with $f_0=5.38$~Hz. Observe that the clipped DWPT load introduces harmonics of lower amplitude compared to the load generated by the multiplicative model (see~\cite{gupta2025dynamicmodelingloaddemand} for more details on the comparison between different DWPT load models).}
\label{fig:trapezoid}
\end{figure}

The signal $h_T(t)$ is a train of trapezoidal pulses of period $T$; it is a scaled version of the blue waveform shown in Fig.~\ref{fig:trapezoid}. Although $h_T(t)$ has infinite duration, it is a good approximation of the finite-duration signal $p_n(t)$ because the number of coil segments $K$ is large. The signal $h_T(t)$ has been time-shifted, so it is even and symmetric about $t=0$. 

Given that $h_T(t)$ is periodic, it is amenable to the Fourier series (FS) expansion
\begin{equation}\label{eq:h(t)2}
h_T(t)=\sum_{m=-\infty}^{\infty}c_m e^{jm\omega_0 t} \,,
\end{equation}
where $\omega_0 = {2\pi}/{T}={2\pi v}/{D}$ is the fundamental frequency and $c_m$ is the FS coefficient related to the $m$-th harmonic. Given that $h_T(t)$ is real-valued and even-symmetric, the FS coefficients are real-valued and even-symmetric (see Fig.~\ref{fig:trapezoid}):
\begin{equation}\label{eq:fscoeffs}
c_m=
\begin{cases}
    \frac{\ell_T\ell_R}{D},&m=0\\
    \frac{D}{m^2 \pi^2} \sin\left(m\pi\frac{\ell_R}{D}\right) \sin\left(m\pi\frac{\ell_T}{D}\right), &m\neq 0.
\end{cases}
\end{equation}
For the INDOT testbed, we get $c_5=0$ because $\ell_T/D=4/5$. 

Based on~\eqref{eq:ConstantSpeedpnt2} and the time-shifting property of the FS, the FS expansion of $p_n(t)$ is 
\begin{equation}\label{eq:pnfs}
p_n(t) = a_n h_T(t-t_n)=a_n \sum_{m=-\infty}^{\infty} c_m e^{jm\omega_0t}e^{-jm\omega_0t_n}.
\end{equation}
The timing $t_n$ affects $p_n(t)$ through harmonic trigonometric functions with fundamental period $T$, and thus introduces a phase shift on the periodic signal $h_T(t)$. To simplify the presentation, we will henceforth slightly abuse notation and use $t_n$ to denote the original timing of EV$_n$ modulo $T$. Therefore, the timing $t_n\in [0,T)$ now captures the relative timing at which EV$_n$ aligns with the pattern of Tx coils. 

From \eqref{eq:h(t)2}--\eqref{eq:pnfs}, it is worth identifying the following properties of $h_T(t)$ and $p_n(t)$:
\begin{enumerate}
\item[\emph{p1)}] The fundamental frequency $\omega_0=2\pi v/D$ grows with the EV speed and the inverse of~$D$. For the INDOT testbed, we get $f_0=5.38$~Hz for $v=24.56$~m/s ($55$~mph), and $f_0$ falls below 2~Hz when $v<9.83$~m/s (22~mph).
\item[\emph{p2)}] Although the locations of harmonics $m\omega_0$ depend on the EV speed, the FS coefficients $c_m$ depend only on Rx/Tx dimensions. Applying the definition of THC to the periodic load $p_n(t)$ in \eqref{eq:pnfs} leads to the alternative THC expression
\begin{equation}\label{eq:thc}
\text{THC} = \frac{\sqrt{2\sum_{m=1}^{\infty}|c_m|^2}}{c_0}\times 100\%.
\end{equation} 
For the INDOT testbed, the signal $h_T(t)$ has $\text{THC}_h=26\%$. The THC should not be confused with the \emph{total harmonic distortion} (THD), which measures the relative power contained in the non-fundamental frequencies of an AC electric signal relative to its fundamental frequency.
\item[\emph{p3)}] Only the first few harmonics contribute notably to $p_n(t)$. For the INDOT ER testbed, with $c_1=0.18c_0$, the first harmonic contributes 25.45\% to the THC of 26\%. 
\end{enumerate}

Having characterized the frequency content of $h_T(t)$ and $p_n(t)$, we shift focus to the DWPT load $p(t)$. Plugging \eqref{eq:pnfs} into \eqref{eq:total} shows that $p(t)$ is also periodic: 
\begin{equation}\label{eq:FS of p(t)}
p(t) = \sum_{n=1}^N p_n(t)=\sum_{m=-\infty}^{ \infty} \tilde{c}_m e^{jm\omega_ot}
\end{equation}
with fundamental frequency $\omega_0={2\pi v}/{D}$ and FS coefficients
\begin{equation}\label{eq:cm_tilde}
\tilde{c}_m = c_m\sum_{n=1}^N a_n e^{-jm\omega_0t_n}.
\end{equation}
This leads to the next obvious, yet important, remark.

\begin{remark}\label{re:v/d}
Under constant EV speed, the DWPT load $p(t)$ can include frequency components below 2~Hz (that is, within the range of inter-area oscillations) if the EV speed $v$ in m/s and the coil segment length $D$ in m satisfy $v<2D$.
\end{remark}

We next study the frequency content of the DWPT load under the first three traffic scenarios. 

\subsection{Scenario S$_1$: Perfect Time Synchronization}\label{subsec:s1}
According to \eqref{eq:cm_tilde}, the FS coefficients $\tilde{c}_m$ of $p(t)$ critically depend on EV timings. This subsection considers the worst-case scenario in terms of THC of $p(t)$. To build up some intuition, suppose EV densities are equal for now, that is, $a_n=a$ for all $n$. If EVs have equal timings, EVs drive over the coil segments in a synchronized fashion. In this case, EV loads $p_n(t)$ add up constructively, and the total load $p(t)$ becomes a time-shifted version of $Nah_T(t)$. Then, the total DWPT load inherits properties \emph{p1)}--\emph{p3)}. For general EV densities $a_n$, the ensuing lemma identifies worst-case EV settings that maximize THC; all proofs are provided in the appendix.

\begin{lemma}\label{le:s1}
Under the assumption of a common constant EV speed, the maximum THC is realized if EVs have equal timings $\{t_n\}_{n=1}^N$, regardless of the EV densities $\{a_n\}_{n=1}^N$. Then, the maximum THC for $p(t)$ coincides with the THC of $h_T(t)$. Equal timings also maximize the harmonic ratios $|\tilde{c}_m|/\tilde{c}_0$ for $m\geq 1$. If $a_n=a$ for all $n$, the Fourier transform of $p(t)$ is
\[P(\omega)=2\pi Na\sum_{m=-\infty}^{\infty}c_m\delta(\omega-m\omega_0) \, ,\]
where $\delta(\omega)$ is the Dirac delta function.
\end{lemma}

Lemma~\ref{le:s1} suggests that if EVs move synchronously and their spacings align with the ER coil spacing, the DWPT load spectrum can have THC equal to that of $h_T(t)$, which is 26\% for the INDOT testbed. Both the DC and harmonic components of $p(t)$ then scale with $N$. Moreover, as Remark~\ref{re:v/d} suggests, single-frequency components can occur in the 0--2~Hz range at slower EV speeds. 

Admittedly, EVs are unlikely to be perfectly synchronized in practice. EVs may be approximately synchronized under heavy traffic conditions, when drivers may generally follow the \emph{2-second rule}. According to this rule, a driver maintains a two-second spacing between their vehicle and the leading vehicle. If the distance $2v$ traveled in two seconds is an integer multiple of $D$, EVs can get synchronized. Assuming \emph{S$_1$} as the worst case, we next study more realistic traffic conditions.

\subsection{Scenario S$_2$: Independent Uniformly Distributed Timings}\label{subsec:s2}
Given that individual EV loads can be rather arbitrary, any deterministic modeling of $\{p_n(t)\}_{n=1}^N$ may be of limited value. For a more systematic characterization of the DWPT load spectrum, we suggest treating these signals as stochastic. To find the frequency content of $p(t)$, one may apply the Fourier transform to $p(t)$. Unfortunately, the Fourier transform is undefined for stochastic signals. Instead, one may define the \emph{power spectral density} (PSD) $S_p(\omega)$ of $p(t)$. The PSD describes how the power of a stochastic signal is distributed across frequencies. Note that the term \emph{power} is used here in two contexts. The signal $p(t)$ denotes the electric power consumed by the ER and is measured in W (Watts), but we can also define the power of the signal $p(t)$, measured in W$^2$.

A sufficient condition for the PSD of a stochastic signal to exist is that the signal is \emph{wide sense stationary} (WSS)~\cite[Ch.~4]{proakis}. A stochastic signal $p(t)$ is WSS if its mean does not vary with time and its autocorrelation depends only on the lag:
\begin{align*}
\mathbb{E}[p(t)]=\mu_p,\quad \text{and} \quad \mathbb{E}[p(t)p(t+\tau)]=R_p(\tau),\quad \forall t.
\end{align*}
If $p(t)$ is WSS, its PSD can be computed as the Fourier transform of its autocorrelation function as
\begin{equation}\label{eq:psd}
S_p(\omega)=\mcF[R_p(\tau)].
\end{equation}
We next compute the PSD of $p(t)$ under different statistical models on EV parameters $\{(a_n,t_n)\}_{n=1}^N$. 

Section~\ref{subsec:s1} considered the worst-case scenario where EVs are perfectly synchronized. This subsection examines the opposite extreme: no synchronization among EVs. This scenario, termed \emph{S$_2$}, can occur when EVs are spaced sufficiently far apart and move freely. Under \emph{S$_2$}, we model EV timings $t_n$ as independent and identically distributed (i.i.d.) random variables uniformly drawn from $[0,T)$. It is reasonable to assume that $t_n$ is uniformly distributed through any time interval. We select the interval $[0,T)$ without loss of generality for mathematical convenience. We model EV densities $a_n$ as i.i.d. random variables of given mean and variance. It is reasonable to assume that EV densities are independent of EV timings.

As $p(t)$ is the sum of several randomly time-shifted replicas of $h_T(t)$, one may conjecture that $p(t)$ approaches a constant (time-invariant) signal as $N$ grows, so its harmonics eventually disappear. The resulting expression establishes the correct scaling law for DWPT harmonics. 

\begin{lemma}\label{le:s2}
Suppose EVs move at constant speed $v$; EV densities $a_n$ are i.i.d. with mean $\mu_a$ and variance $\sigma_a^2$; and EV timings $t_n$ are i.i.d. uniformly within $[0,T)$. Then, the DWPT load $p(t)$ is WSS with mean $\mu_p(t) =N\mu_a c_0$ and PSD:
\begin{align*}
S_p(\omega)&=2\pi (N^2\mu_a^2 + N \sigma_a^2)c_0^2\delta(\omega)\\
&\quad  + 2\pi N(\mu_a^2+\sigma_a^2)\sum_{m\neq 0} c_{m}^2 \delta(\omega-m\omega_0).
\end{align*}
\end{lemma}

According to Lemma~\ref{le:s2}, the spectrum of $p(t)$ under \emph{S$_2$} contains the same harmonics as in \emph{S$_1$}. Recall that under \emph{S$_1$}, the DC term and harmonics of $p(t)$ scale with $N$. Under \emph{S$_2$}, harmonics do \emph{not} disappear, but scale differently. To simplify the analysis, suppose $\sigma_a^2=0$ and $\mu_a=a$, so EV densities become deterministic and conform with Corollary~\ref{le:s1}. Then, Lemma~\ref{le:s2} explains that the DC term under \emph{S$_2$} scales with $Nac_0$, whereas its $m$-th harmonic scales with $\sqrt{N}a|c_m|$. The key conclusion from this analysis is that even when EV timings are independent, DWPT harmonics do not disappear, but scale with $\sqrt{N}$ rather than $N$.

For stochastic signals $p(t)$, the THC in~\eqref{eq:THC_og} can be expressed in stochastic terms as
\begin{equation}\label{eq:THC_cont}
\text{THC}_p = \sqrt{\frac{2\int_{0}^{\infty} S_p(\omega)~d\omega  - \lim_{\epsilon\rightarrow 0} \int_{-\epsilon}^{\epsilon} S_p(\omega)~d\omega }{\lim_{\epsilon\rightarrow 0} \int_{-\epsilon}^{\epsilon} S_p(\omega)~d\omega }}\times 100\%.
\end{equation}
For periodic stochastic signals with FS coefficients $\tilde{c}_m$, the THC formula of \eqref{eq:THC_cont} simplifies as 
\begin{equation}\label{eq:THC_stochastic}
    \text{THC}_p= \sqrt{\frac{2\sum_{m=1}^{\infty} \expect [|\tilde{c}_m|^2]}{\expect [|\tilde{c}_0|^2]}}, 
\end{equation}
where $\expect [|\tilde{c}_m|^2]$ is the magnitude of PSD $S_p(\omega)$ at frequency $m\omega_0$. Under the \emph{S$_2$} setting of unsynchronized traffic flow, Lemma~\ref{le:s2} shows that for the special case of $\sigma_a = 0$, the THC of the total DWPT load decreases with $\sqrt{N}$ as
\[ \text{THC}_p = \text{THC}_h/\sqrt{N}.\]
Therefore, the DWPT load may have insignificant frequency content as $N$ grows, as verified numerically in Sections~\ref{sec:matlab_tests}--\ref{sec:sumo}. 

\subsection{Scenario S$_3$: Vehicle Platoons}\label{subsec:s3}
Vehicles driving on a highway often form \emph{platoons}, i.e., groups of vehicles moving together at the same speed. This subsection studies the effect of EV platoons on the DWPT signal. Suppose the $N$ EVs served by an ER segment form platoons. To ease the analysis under \emph{S$_3$}, we assume: 
\begin{itemize}
    \item[\emph{i)}] Each platoon has the same number of EVs. {The number of EVs per platoon will be denoted by $Q$};
    \item[\emph{ii)}] EVs within a platoon are perfectly time-synchronized;
    \item[\emph{iii)}] The timing for each platoon is i.i.d. and uniformly in $[0,T)$;
    \item[\emph{iv)}] EV densities are i.i.d. of given mean $\mu_a$ and variance $\sigma_a^2$.
\end{itemize}
Assumption \emph{ii)} considers the worst-case scenario where EVs within a platoon move synchronized and at spacings that are integer multiples of $D$. Scenario \emph{S$_3$} can be studied by combining \emph{S$_1$} and \emph{S$_2$}. If EVs within a platoon are synchronized, they can be modeled as a single EV with a higher rating. On the other hand, timings across platoons can be modeled as independent and uniformly distributed. Based on these two points, we next derive the PSD of $p(t)$ under \emph{S$_3$}.

\begin{lemma}\label{le:s3}
Under the assumptions of {S$_3$}, the mean of $p(t)$ is $N\mu_a c_0$ and its PSD is
\begin{align*}
S_p(\omega)&=2\pi (N^2\mu_a^2 + N \sigma_a^2)c_0^2\delta(\omega)\\
&\quad  + 2\pi N\left(Q\mu_a^2+\sigma_a^2\right)\sum_{m\neq 0} c_{m}^2 \delta(\omega-m\omega_0).
\end{align*}
\end{lemma} 

To quantify the effect of platoon formations, compare Lemma~\ref{le:s3} to Lemma~\ref{le:s2}. To ease the comparison, consider deterministic EV densities by setting $\sigma_a =0$ and $\mu_a=a$. Although the DC component under \emph{S$_3$} remains $Nac_0$, harmonics scale now as $\sqrt{NQ}a|c_m|$ instead of $\sqrt{N}a|c_m|$. Clearly, harmonic components are more prominent in \emph{S$_3$} compared to \emph{S$_2$}. This is expected as platoon formations imply less smoothing across time-shifted replicas of $h_T(t)$. Therefore, platoon formations can increase the THC in $p(t)$. When $Q=1$, scenario \emph{S$_3$} becomes equivalent to \emph{S$_2$}. On the other extreme, when $Q=N$, scenario \emph{S$_3$} becomes \emph{S$_1$}.

\begin{remark}\label{re:EVnumber}
Our analysis assumes a constant number $N$ of EVs on the ER segment. A more detailed analysis should model $N$ as a random variable. Traffic flow studies usually postulate that EV arrivals follow a Poisson distribution parameterized by the \emph{mean arrival rate} $\lambda_a$, measured in EV/s. If all EVs move at constant speed $v$, the total number $N$ of EVs driving on an ER segment of length $L$ follows a Poisson distribution with mean $\lambda_N = \lambda_a L/v$;  see~\cite[Sec~8.3]{Ross}. If $\lambda_a=0.3$~EV/s, $L=10$~mi, and $v=55$~mph, we get the mean value of $\lambda_N=192$. Moreover, the probability of $N$ being within $\pm 10\%$ of its mean value $\lambda_N=192$ is 84\%. To simplify the analysis, we treat $N$ as deterministic and set it to its mean value $\lambda_N$ determined by $(\lambda_a,L,v)$.
\end{remark}

\section{EVs Moving at Constant but Unequal Speeds}\label{sec:cons&uneq_vel}
We have hitherto assumed that EVs move at a constant speed, so that the total DWPT load $p(t)$ is periodic. This section considers the scenario where EVs move at constant yet unequal speeds. In this case, although each EV load is periodic, $p(t)$ is aperiodic and thus has a continuous spectrum.

\subsection{Scenario S$_4$: Constant Gaussian-Distributed Speeds}\label{subsec:s4}
Consider first a single EV. Similar to \eqref{eq:pnfs}, if EV$_n$ moves at constant speed $v_n$, its power demand $p_n(t)$ is periodic with fundamental period $T_n=D/v_n$ and fundamental frequency $\omega_n = {2\pi v_n}/{D}$. Signal $p_n(t)$ has the FS expansion:
\begin{equation}\label{eq:FS_unequalvel}
p_n(t) = a_n\sum_{m=-\infty}^{ \infty} c_m e^{jm\omega_n(t-t_n)}
\end{equation} 
with the FS coefficients defined as in \eqref{eq:fscoeffs}. EV densities $a_n$ are again modeled as i.i.d. with mean $\mu_a$ and variance $\sigma_a^2$. Different from Sec.~\ref{sec:cons&equal_vel}, here speeds $v_n$ are modeled as i.i.d. Gaussian random variables distributed as $v_n\sim \mcN(\mu_v,\sigma^2_v)$. If $v_n$ is Gaussian, the fundamental frequency $\omega_n = {2\pi v_n}/{D}$ is also Gaussian with mean and variance:
\[\mu_\omega = \frac{2\pi\mu_v}{D}\quad \text{and}\quad \sigma^2_\omega=\frac{4\pi^2\sigma_v^2}{D^2}.\]
Regarding EV timings $t_n$, it is reasonable to assume they are uniformly distributed in $[0,T_n)$. Parameters $(a_n,v_n,t_n)$ are assumed to be independent across EVs. 

Under \emph{S$_4$}, the total DWPT load can be expressed as
\begin{equation}\label{eq:p(t)_unequalvel}
        p(t) = \sum_{n=1}^N a_n\sum_{m=-\infty}^{ \infty} c_m e^{jm\omega_n(t-t_n)}.
\end{equation} 
Recall that a linear combination of periodic signals is not necessarily periodic. More precisely, signal $p(t)$ is periodic if and only if all $\omega_n$ are integer multiples of a single frequency $\omega_0$. Because this condition is practically impossible, we conclude that $p(t)$ is non-periodic and does not inherit properties \emph{p1)--p3)} discussed in Sec.~\ref{sec:cons&equal_vel}. Even though $p(t)$ does not have an FS expansion, its frequency-domain analysis is still possible if we can determine its PSD. The following lemma establishes that $p(t)$ is WSS and derives its PSD.

\begin{lemma}\label{le:s4}
Under the assumptions of S$_4$, the total DWPT load $p(t)$ is WSS with mean $\mu_p(t) =N\mu_a c_0$ and PSD
\begin{align*}
S_p(\omega)&=2\pi(N^2\mu_a^2 +N\sigma_a^2)c_0^2\delta(\omega)\\
&\quad + 2\pi N(\mu_a^2 + \sigma_a^2) \sum_{m\neq 0} c_m^2 \phi(\omega;m\mu_{\omega},m^2\sigma_{\omega}^2)\, ,
\end{align*}
where $\phi(x;\mu,\sigma^2)$ denotes the Gaussian probability density function (PDF) with mean $\mu$ and variance $\sigma^2$.
\end{lemma}

Lemma~\ref{le:s4} confirms that the PSD of $p(t)$ under \emph{S$_4$} is not discrete, but continuous: It consists of Gaussian-shaped bells or lobes centered around the frequency harmonics $\pm m\mu_{\omega}$. Consider the $m$-th Gaussian lobe. Because its standard deviation scales with $m$, its bandwidth is proportional to $m$, and its peak value is inversely proportional to $m$. Hence, the lobes $\phi(\omega;m\mu_{\omega},m^2\sigma_{\omega}^2)$ reduce in amplitude and expand in bandwidth for higher $m$. 

\emph{Do the Gaussian lobes overlap?} The effective bandwidth of a Gaussian lobe can be shown to be 2.3 times its standard deviation. Here, the effective bandwidth is defined as the difference between two points whose value on the lobe is half the maximum value of the Gaussian lobe. Then, the $m$-th lobe essentially occupies the frequency range:
\[\left[m(\mu_{\omega}- 1.15\sigma_{\omega}),m(\mu_{\omega}+ 1.15\sigma_{\omega})\right]\] 
and the $(m+1)$-th lobe occupies the range:
\[\left[(m+1)(\mu_{\omega}- 1.15\sigma_{\omega}),(m+1)(\mu_{\omega}+ 1.15\sigma_{\omega})\right].\]
For the two lobes not to overlap, we need
\[m(\mu_{\omega}+ 1.15\sigma_{\omega})<(m+1)(\mu_{\omega}- 1.15\sigma_{\omega})\]
which is equivalent to $\mu_{\omega}>1.15(2m+1)\sigma_{\omega}$ or $\mu_v>1.15(2m+1)\sigma_v$. This condition is reasonable to be met for harmonics $m\in\{1,2,3\}$ that are of primary interest. 

The THC for the total DWPT load under S$_4$ can be computed from~\eqref{eq:THC_cont}. According to Lemma~\ref{le:s4}, the DWPT load has precisely the same THC under \emph{S$_2$} and \emph{S$_4$}. This is because the Gaussian PDF integrates to unity over $\omega$. This observation also implies that the energy contributed by the lobe centered around $m\mu_{\omega}$ under \emph{S$_4$} equals the energy of harmonic $m\omega_0$ under \emph{S$_2$}. Interestingly, scenario \emph{S$_2$} can be derived as the limiting case of \emph{S$_4$} when $\sigma_v$ tends to zero and $\mu_v=v$. This follows from the fact that the Dirac delta function is the limiting case of the Gaussian PDF as the variance vanishes.

To ease comparison with previous scenarios, consider again deterministic EV densities by setting $\sigma_a=0$ and $\mu_a=a$. Then, the DC term under \emph{S$_4$} scales with $Nac_0$ and the Gaussian lobe centered around the $m$-th harmonic scales with $\sqrt{N}a|c_m|$. This is consistent because the timings have been assumed to be independent across EVs, as in \emph{S$_2$}. 

Building on the previous lemmas, we next summarize how the THC$_p$ of the DWPT load varies across scenarios.

\begin{corollary}\label{co:compare}
If EVs have identical deterministic densities, the THC$_p$ relates to the THC$_h$ of signal $h_T(t)$ as:
\begin{align*}
\text{\emph{THC}}_p &= \text{\emph{THC}}_h          &&  \text{for S$_1$};\\
\text{\emph{THC}}_p &= \text{\emph{THC}}_h/\sqrt{N} &&  \text{for S$_2$ and S$_4$}; and\\
\text{\emph{THC}}_p &= \text{\emph{THC}}_h\sqrt{Q/N} &&  \text{for S$_3$}.
\end{align*}
\end{corollary}

\begin{remark}\label{re:pes_gm}
This work derives scaling laws for DWPT harmonics using the multiplicative model of DWPT load demand. In our later work~\cite{gupta2025dynamicmodelingloaddemand}, it is established that similar scaling laws apply to DWPT dynamics under the clipping model. This is because clipping only affects FS coefficients, whereas scaling laws are primarily determined by the EV timing distributions. Reference~\cite{gupta2025dynamicmodelingloaddemand} also shows that the clipping model features lesser THC than the multiplicative model. We also show that the harmonics of an EV load decrease in amplitude as the Rx coil length increases. Studying the effect of EV composition PSD on the DWPT load reveals that serving more EVs with longer Rx coils (trucks) does not necessarily result in milder harmonics.
\end{remark}

\section{Periodograms of DWPT Load Demands}\label{sec:matlab_tests}
Sections~\ref{sec:cons&equal_vel} and \ref{sec:cons&uneq_vel} studied the frequency content of the DWPT load under certain traffic models by deriving its PSD $S_p(\omega)$. The PSD is an \emph{ensemble statistic}. It pertains to all possible realizations of the stochastic load $p(t)$. In practice, however, one would observe a single realization of $p(t)$ at a time. Then, naturally, one asks whether the spectrum of any given realization of $p(t)$ is close enough to $S_p(\omega)$.

This question boils down to whether the \emph{periodogram} of a finite-duration realization of $p(t)$ relates to its PSD. This is a well-studied topic in statistical signal processing~\cite{proakis}. Let us briefly review some key points. Consider a DWPT load realization $\hat{p}(t)$ spanning the finite interval $t\in[0,T_p]$ of duration $T_p$. Grid operators typically use windows of 1--5~min to compute spectra of signals in power system studies~\cite{mishra2025understanding}. The periodogram of $\hat{p}(t)$ is computed as the squared magnitude of its Fourier transform as~\cite[Sec~4.3]{proakis}
\begin{equation}\label{eq:periodogram}
\hat{S}_{T_p}(\omega) = \frac{1}{T_p}|\mcF[\hat{p}(t)]|^2. 
\end{equation}
In addition to the definition in \eqref{eq:psd}, the PSD of $p(t)$ can be defined as the mean of periodograms, asymptotically in $T_p$:
\begin{equation}\label{eq:psdvsperiodgram}
S_p(\omega)=\lim_{T_p\rightarrow \infty}\mathbb{E}[\hat{S}_{T_p}(\omega)] \quad \text{for all}~~\omega.
\end{equation}
In plain words, because $\hat{p}(t)$ is random, its periodogram is also random. Each realization $\hat{p}(t)$ yields a different periodogram. The average of periodograms converges to the PSD $S_p(\omega)$ for increasing $T_p$. Hence, the PSD does not coincide with each periodogram. Instead, it describes the average frequency-domain content of the DWPT load. Intuitively, we expect the periodogram to converge to the PSD for longer windows, higher EV arrival rates, and/or longer highway segments. 

In the three tests of this section, we computed numerically the periodograms of 1-min realizations $\hat{p}(t)$, generated according to \emph{S$_2$}--\emph{S$_4$}. More specifically, we simulated the DWPT load consumed by an $L = 16$~km (10~mile)-long ER segment following the specifications of the INDOT ER testbed. EV loads were computed as trapezoidal pulse trains. EV power densities $a_n$ were simulated as i.i.d. Gaussian random variables with mean $\mu_a=94.45$~kW/m and standard deviation $\sigma_a=5.55$~kW/m; see~\cite{Aaron22}. For \emph{S$_2$} and \emph{S$_3$}, speed was set to $v=24.56$~m/s. At this speed, an EV takes about 10~min to cross the ER segment.

\begin{figure}[t]
\centering    
\includegraphics[width=\linewidth]{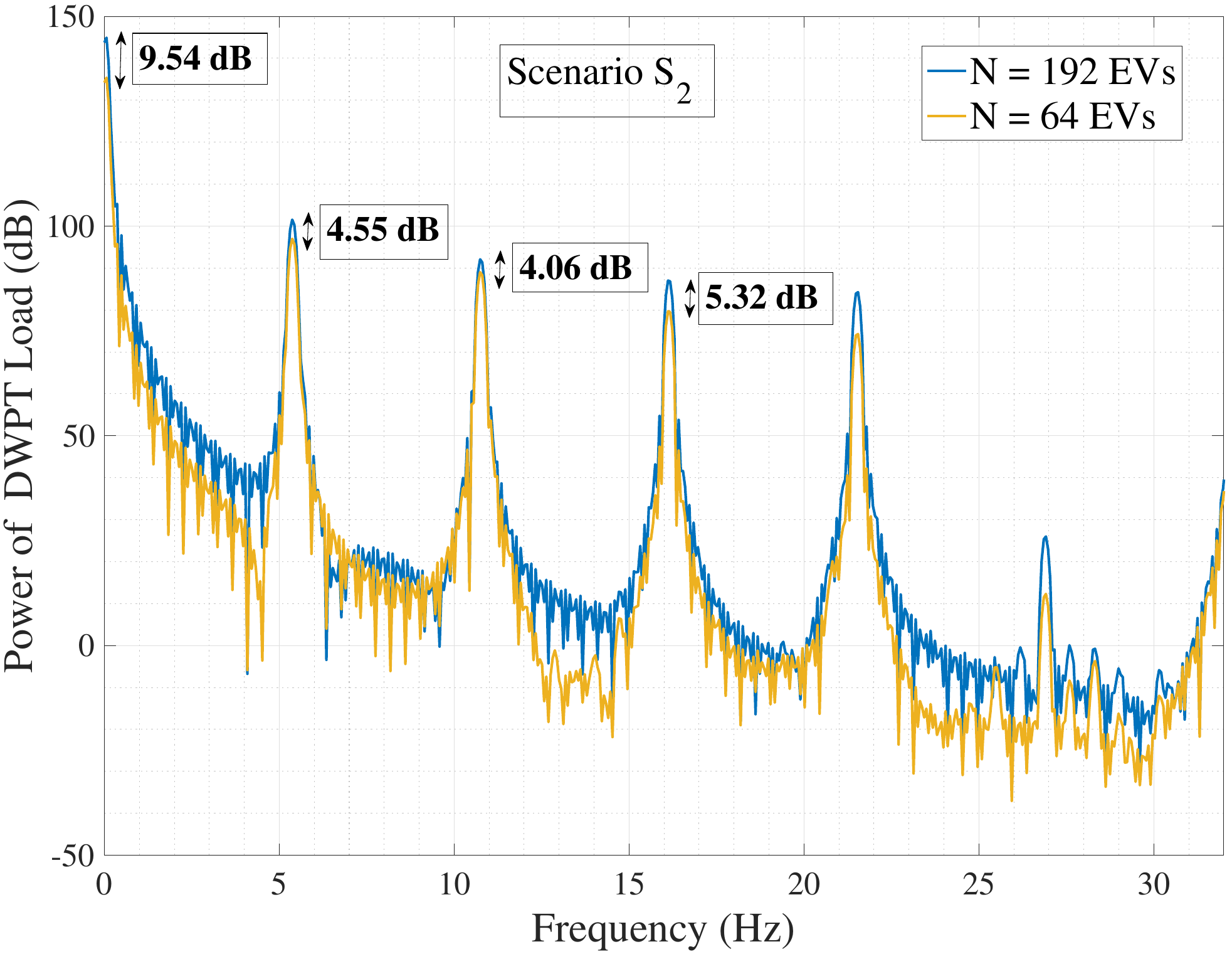}
\caption{Periodograms of 1-min DWPT loads simulated under scenario \emph{S$_2$} for two values of $N$. The magnitude and location of harmonics agree with those predicted by Lemma~\ref{le:s2}. Moreover, as asserted by Lemma~\ref{le:s2}, the ratio of the two periodograms should be $10\log_{10}(192/64)^2 = 9.54$~dB at DC, and about $10\log_{10}\left({192}/{64}\right) = 4.77$~dB at harmonic components.}
    \label{fig:S2_analytical_psd}
\end{figure}

\emph{Periodogram under S$_2$.} To compute the periodogram under \emph{S$_2$}, EV timings were independently uniformly drawn from $[0,T)$, where $T=D/v$. Because independent timings are likely to occur under lighter traffic, we considered two relatively low mean arrival rates, $\lambda_a=0.1$ and $0.3$~EV/s, yielding $N=64$ and $192$~EVs, respectively. Figure~\ref{fig:S2_analytical_psd} depicts the periodogram of a single 1-min realization of $\hat{p}(t)$ after applying a Hanning window. This periodogram differs from the PSD of Lemma~\ref{le:s2} as it relies on a single, finite-duration realization $\hat{p}(t)$. Nonetheless, the locations of harmonic peaks coincide with those predicted by the analysis. For $N=64$ and $N=192$, the DC component of $\hat{p}(t)$ was $8.70$~MW and $26.13$~MW, respectively. Integrating the two periodograms around the first harmonic yielded $357$~kW and $680$~kW for the two values of $N$, respectively. This verifies that although the DC component scales with $N$, harmonics scale with $\sqrt{N}$. In addition, we evaluated the THC for multiple realizations $\hat{p}(t)$ using~\eqref{eq:THC_cont} and found it to be 3--6\% and 1--3\% for the two values of $N$, indicating that the THC decreases with $N$.


\begin{figure}[t]
\centering
\includegraphics[width=\linewidth]{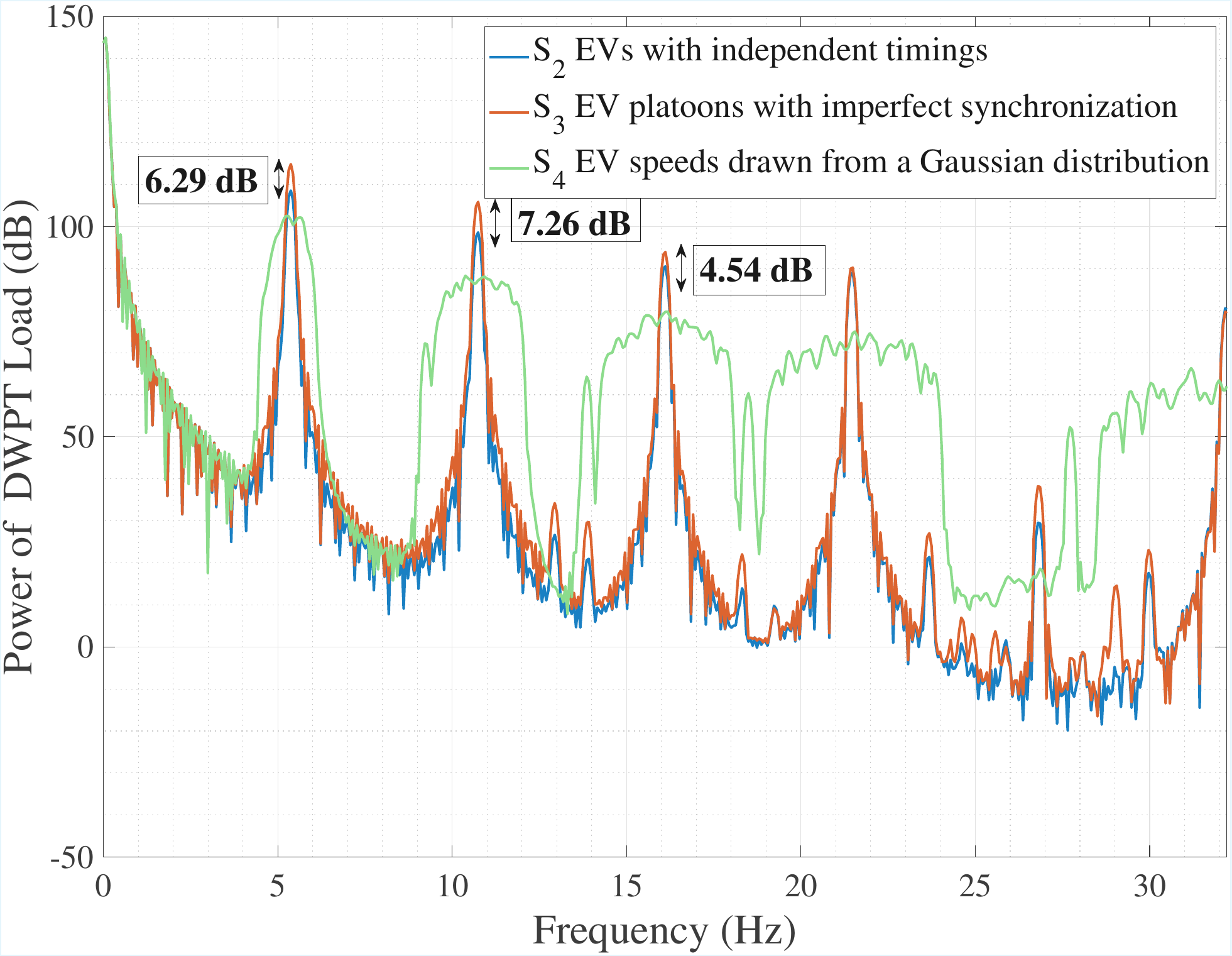}
    \caption{Periodograms of 1-min DWPT loads simulated under \emph{S$_2$}--\emph{S$_4$} for $N=192$ EVs. Harmonics increase in power under \emph{S$_3$} due to platoons. The increase is about $10\log_{10}Q = 6.98$~dB per Lemmas~\ref{le:s2}--\ref{le:s3}. The lobes under \emph{S$_4$} are centered around the harmonics of \emph{S$_2$} because $\mu_v = 55$~mph. lobes reduce in magnitude and spread in bandwidth with $m$ per Lemma~\ref{le:s4}.}
    \label{fig:S234_analytical_psd}
\end{figure}

\emph{Periodogram under S$_3$.} In this case, we simulated $N=192$ EVs moving in platoons of approximately $Q=5$ EVs per platoon. To capture imperfect synchronization, we first sampled the timing of each platoon uniformly from $[0,T)$ independently. Then, for each EV within a platoon, we added an additional time shift, uniformly and independently drawn from $[0,T/5]$. Figure~\ref{fig:S234_analytical_psd} illustrates the obtained periodogram. Although EVs within a platoon are not perfectly synchronized, the obtained spectrum contains higher harmonic content compared to the spectrum under \emph{S$_2$}. As dictated by Lemmas \ref{le:s2} and \ref{le:s3}, although the DC components are identical (barring the deviations of a single periodogram from its mean), harmonics do increase in power under \emph{S$_3$}, as they now scale as $\sqrt{QN}$ instead of $\sqrt{N}$. In addition, we computed the THC across different realizations and found it to be around $3\%$ with imperfect synchronization and $16\%$ with perfect synchronization within platoons.  

\emph{Periodogram under S$_4$.} To simulate \emph{S$_4$}, EV speeds $v_n$ were drawn independently from a Gaussian PDF with $\mu_v =24.56$~m/s ($55$~mph) and $\sigma_v=1.23$~m/s ($2.75$~mph), while EV timings $t_n$ were i.i.d. and uniform in $[0,D/v_n)$. Figure~\ref{fig:S234_analytical_psd} compares the periodograms obtained under \emph{S$_4$} to \emph{S$_2$} and \emph{S$_3$} for $N=192$ EVs. Unequal speeds cause spectrum spreading under \emph{S$_4$}. As asserted by Lemma~\ref{le:s4}, the bandwidth of the lobes increases with $m$. The magnitude of the peaks is smaller under \emph{S$_4$} rather than \emph{S$_2$}, as expected. The test also corroborates that the THCs for \emph{S$_2$} and \emph{S$_4$} are equal on average.  In theory, the first $8$ lobes should not overlap as the condition $\mu_v>1.15(2m+1)\sigma_v$ is satisfied. Some overlap is observed here due to windowing and finite-sample effects. 

\section{EVs Moving at Time-Varying Speeds}\label{sec:time_varying}
This section studies the spectrum of individual EVs moving at time-varying speeds. 
\subsection{Scenario S$_5$: EV Moving at Time-Varying Speed}
The instantaneous speed of EV$_n$ can be decomposed as 
\[ v_n(t) = \bar{v}_n + \tilde{v}_n(t)\]
where $\tilde{v}_n(t)$ is the deviation of the instantaneous speed from its time-averaged speed $\bar{v}_n$. Then, the position of EV$_n$ at time $t$ can be expressed as
\begin{equation}\label{eq:xn(t)}
    x_n(t) = \bar{v}_nt + \int_{0}^t \tilde{v}_n(\tau) d\tau.
\end{equation}
According to~\eqref{eq:pn(x)}, the power drawn by EV$_n$ is a periodic function of $x_n$ and therefore is amenable to the FS expansion
\begin{equation}\label{eq:p_space}
p_n(t) = \bar{p}_n(x_n(t)) = a_n \sum_{m=-\infty}^{\infty} c_m e^{jm\omega_s x_n(t)}, 
\end{equation}
where $\omega_s = 2\pi/D$ is the \emph{spatial frequency} . Substituting~\eqref{eq:xn(t)} into \eqref{eq:p_space} yields
\begin{equation}\label{eq:pn_FM}
    p_n(t) = a_n \sum_{m=-\infty}^{\infty} c_m e^{jm\omega_s \bar{v}_n t} \cdot e^{jm\omega_s \int_{0}^t \tilde{v}_n(\tau) d\tau}.
\end{equation}

For general aperiodic signals $\tilde{v}_n(t)$, it is difficult to characterize the spectrum of $p_n(t)$. However, if $\tilde{v}_n(t)$ is approximated as periodic with frequency $\omega_n = 2\pi/T_n$, the second complex exponential in \eqref{eq:pn_FM} is also amenable to an FS expansion as~\cite[Sec 3.3.2]{proakis}:
\begin{equation}\label{eq:periodic approximation}
e^{jm\omega_s \int_{0}^t \tilde{v}_n(\tau) d\tau} = \sum_{k=-\infty}^{\infty} \beta_k(m) e^{jk\omega_n t}, 
\end{equation}
where $\beta_k(m)$ denotes the $k$-th FS coefficient corresponding to the $m$-th harmonic. Reference~\cite[Sec 3.3.2]{proakis} computes these coefficients for $\tilde{v}_n(t)$ being a sinusoidal signal. Although there is no closed-form expression for $\beta_k(m)$ for general periodic $\tilde{v}_n(t)$, the coefficients $\beta_k(m)$ can be shown to decrease exponentially with $k$. Substituting \eqref{eq:periodic approximation} into~\eqref{eq:pn_FM} provides
\begin{equation}
    p_n(t) = a_n \sum_{m=-\infty}^{\infty}\sum_{k=-\infty}^{\infty} c_m \beta_k(m) e^{j(m\omega_s \bar{v}_n+k\omega_n)t}. 
\end{equation}
Therefore, the spectrum of $p_n(t)$ consists of peaks at frequencies $m\omega_s \bar{v}_n + k \omega_n$ for all $k$ and $m$. However, the peaks corresponding to larger $k$'s have low amplitude as the coefficients $\beta_k(m)$ decrease in $k$. This special case shows the technical difficulty in characterizing the spectrum of a single EV load, let alone an aggregation of them. However, the spectral spreading due to time-varying speeds can be observed numerically by computing spectra of DWPT loads derived from real-world traffic tests in Section~\ref{sec:sumo}.

\section{Tests Using the SUMO Simulator}\label{sec:sumo}
To better understand the spectra of real-world traffic, we simulated DWPT loads with the help of SUMO~\cite{SUMO18}, an open-source microscopic traffic flow simulator. It can generate realistic vehicle trajectories on a roadway under specified conditions. Given the trajectory of each EV$_n$, its load $p_n(t)$ can be computed using \eqref{eq:pn(x)} and \eqref{eq:pn(t)}. The demand $p(t)$ is computed as the sum of EV loads. 

We simulated an $L=4$~km ($2.5$~mile) ER segment with two lanes. The DWPT lane is the rightmost one. Vehicles were allowed to change lanes to overtake the preceding vehicle, if needed, by temporarily moving to the non-DWPT leftmost lane. All vehicles driving in the DWPT lane were considered to be EVs of the same type (class-8 trucks). The DWPT parameters $(\ell_T,\ell_R,d)$ matched the INDOT testbed specifications. EV power densities were drawn independently from a Gaussian distribution with mean $\mu_a=94.45$~kW/m and standard deviation $\sigma_a=5.55$~kW/m. For SUMO parameters, we set the maximum vehicle speed to be $36$~m/s ($80$~mph), while vehicle acceleration was constrained to lie within $[-4.5,2.6]~\text{m/s}^2$. Successive vehicles were spaced by more than $2.5$~m and $1$~sec. Traffic followed the Krauss car-following model~\cite{Krauss98}. According to this model, the driver of vehicle $n$ has a \emph{desired speed} $v_n^d$, but may have to slow down depending on the speed of the leading vehicle.

\begin{figure}[t]
\centering
\includegraphics[width=\linewidth]{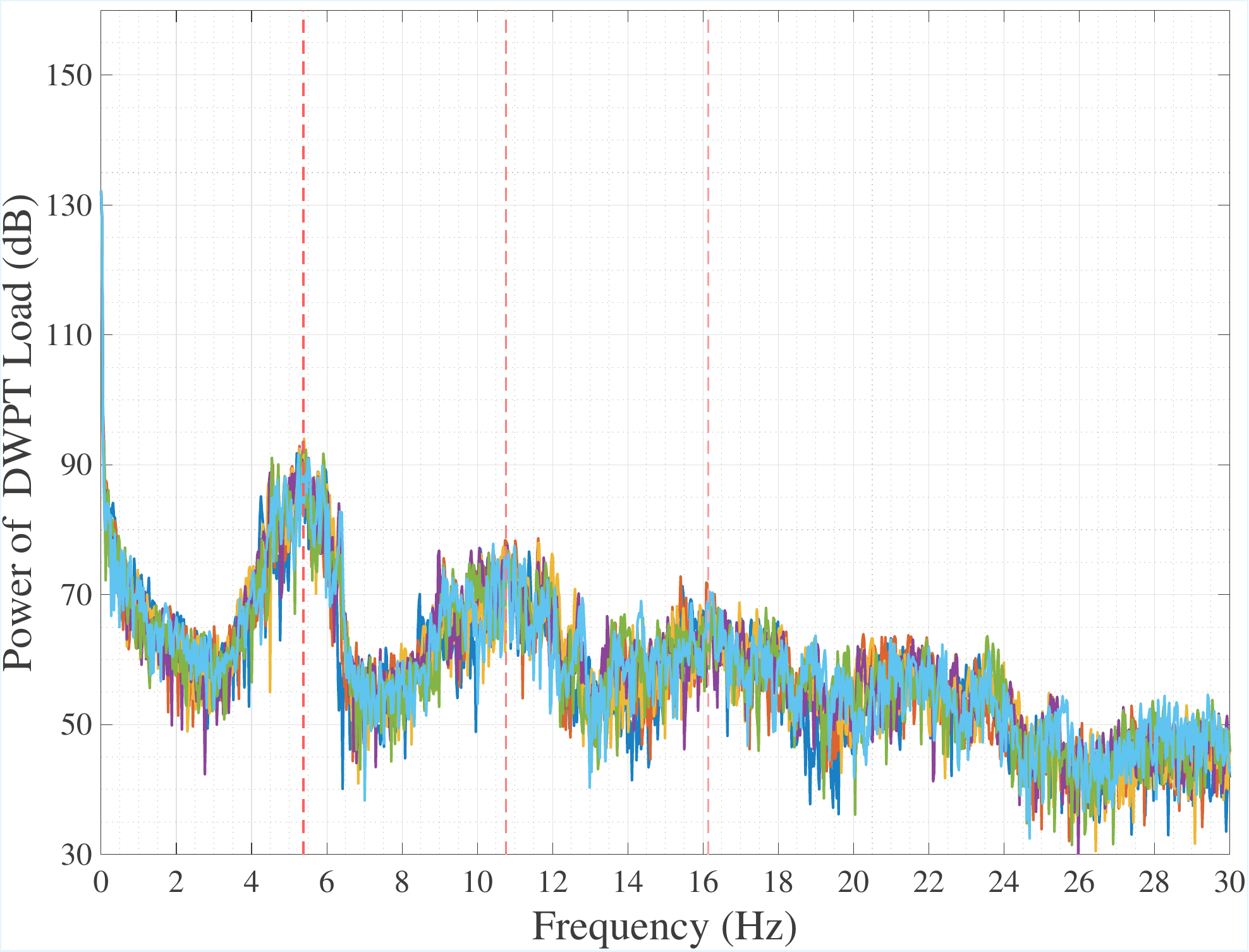}
\caption{{Periodograms computed from 1-min DWPT loads simulated using SUMO under light-traffic conditions. The periodograms contain lobes centered around the harmonics of $f_0=5.38$~Hz associated with mean speed $\mu_v=24.56$~m/s ($55$~mph) and the INDOT testbed segment spacing. lobes decay in magnitude and spread in bandwidth with increasing $m$, as asserted by Lemma~\ref{le:s4}. Periodograms across realizations are similar in shape. lobes start overlapping beyond the third harmonic.}}
\label{fig:sumo_psd}
\end{figure}

\emph{Free-flowing traffic conditions.} With our first test using SUMO, we wanted to test whether \emph{S$_4$} can explain the spectrum of the DWPT load under light traffic. {Additionally, we demonstrate the spreading of the spectra due to EVs moving with time-varying speeds as highlighted in $\emph{S$_5$}$.} To this end, we simulated a mean arrival rate of $\lambda_a = 0.21$~EV/s. The desired EV speeds $v_n^d$ were drawn from a truncated Gaussian distribution with mean $\mu_v =24.56$~m/s, standard deviation $\sigma_v = 2.45$~m/s ($5.5$~mph), while the lower and upper cutoff speeds were set to $20.11$~m/s ($45$~mph) and $29.05$~m/s ($65$~mph), respectively. We simulated traffic for over 40~min, and then randomly selected 1-min intervals as $\hat{p}(t)$. The total number of EVs driving on the ER segment remained relatively constant at $N = 49$~EVs {with a maximum variation of $7\%$} across all 1-min intervals; see Remark~\ref{re:EVnumber}. Figure~\ref{fig:sumo_psd} depicts the periodograms obtained over 6 Monte Carlo runs. The DC component is $6.8$~MW, the power contained around the first harmonic is $613.70$~kW. THC in the range of 6--7\% was observed for different $\hat{p}(t)$ segments.

Because vehicles move under free-flow traffic conditions, their actual speeds are close to their respective desired speeds $v_n^d$. Therefore, for all 1-min realizations $\hat{p}(t)$, EV instantaneous speeds can roughly be described by the aforementioned Gaussian distribution. Because EV average speeds $\bar{v}_n$ are close to $v_n^d$, the spectrum contains main lobes at frequencies $\mu_v/D$. Additional sidelobes appear as EVs move with time-varying speeds as postulated in \emph{S$_5$}. Overall, the key features of the spectra in Fig.~\ref{fig:sumo_psd} agree with the analysis under \emph{S$_4$} and \emph{S$_5$}. 

\begin{figure}[t]
\centering
\includegraphics[width=\linewidth]{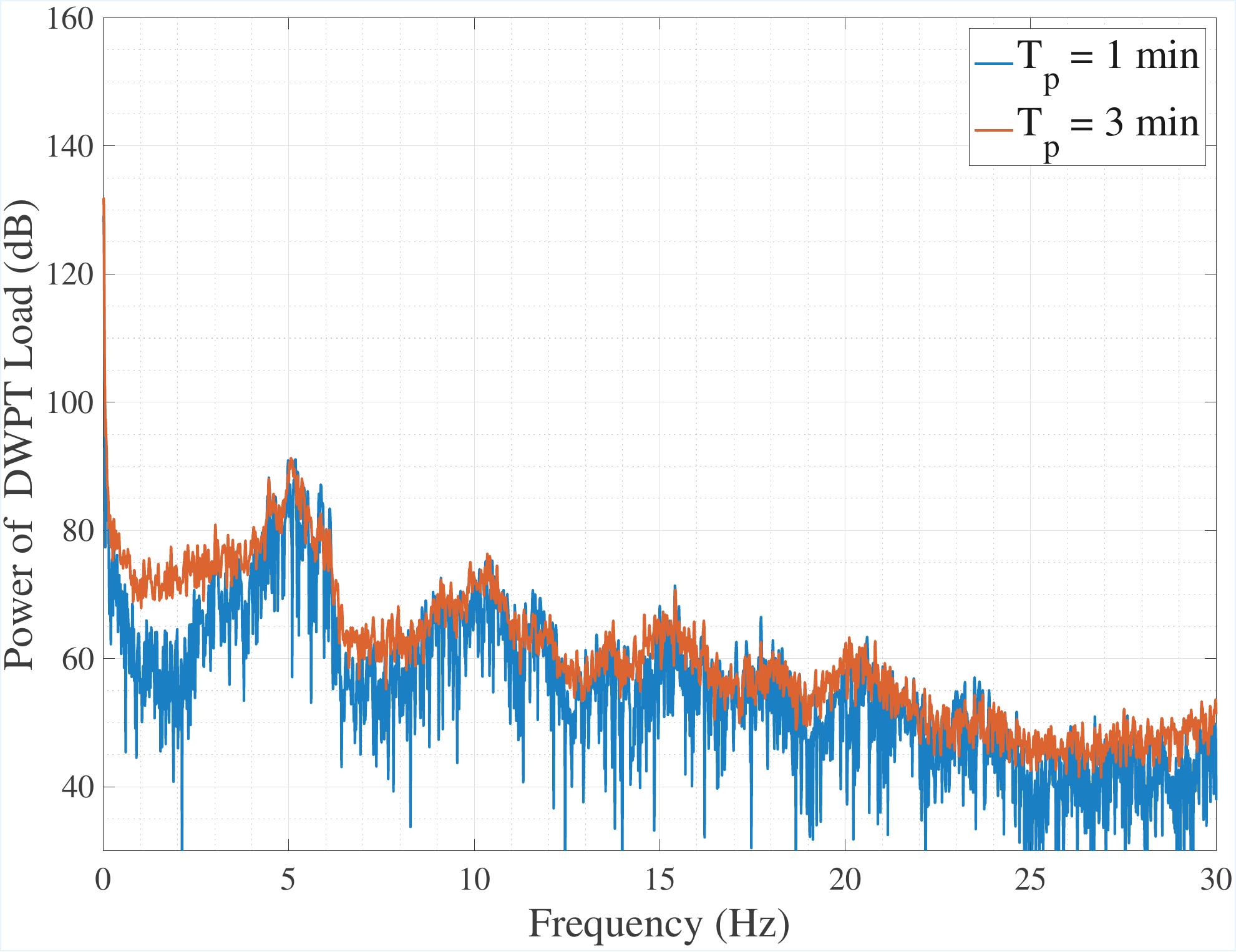}
\caption{DWPT periodograms computed over two different window durations $T_p$. Longer windows enhance frequency resolution as the number of time samples increases.}
\label{fig:windowing}
\end{figure}

The next test studies how the signal duration $T_p$ affects the periodogram $\hat{S}_p(\omega)$. As explained earlier, the periodogram is expected to converge to the PSD when the signal $\hat{p}(t)$ is of infinite duration. To that end, we compare the frequency content of the power drawn by EVs under free-flowing traffic conditions by evaluating the PSD of DWPT loads of different time resolution (See Fig.~\ref{fig:windowing}). We observe that increasing the signal duration leads to sharper peaks. This is because a longer time duration enhances frequency resolution.

\begin{figure}[t]
\centering
\includegraphics[width=\linewidth]{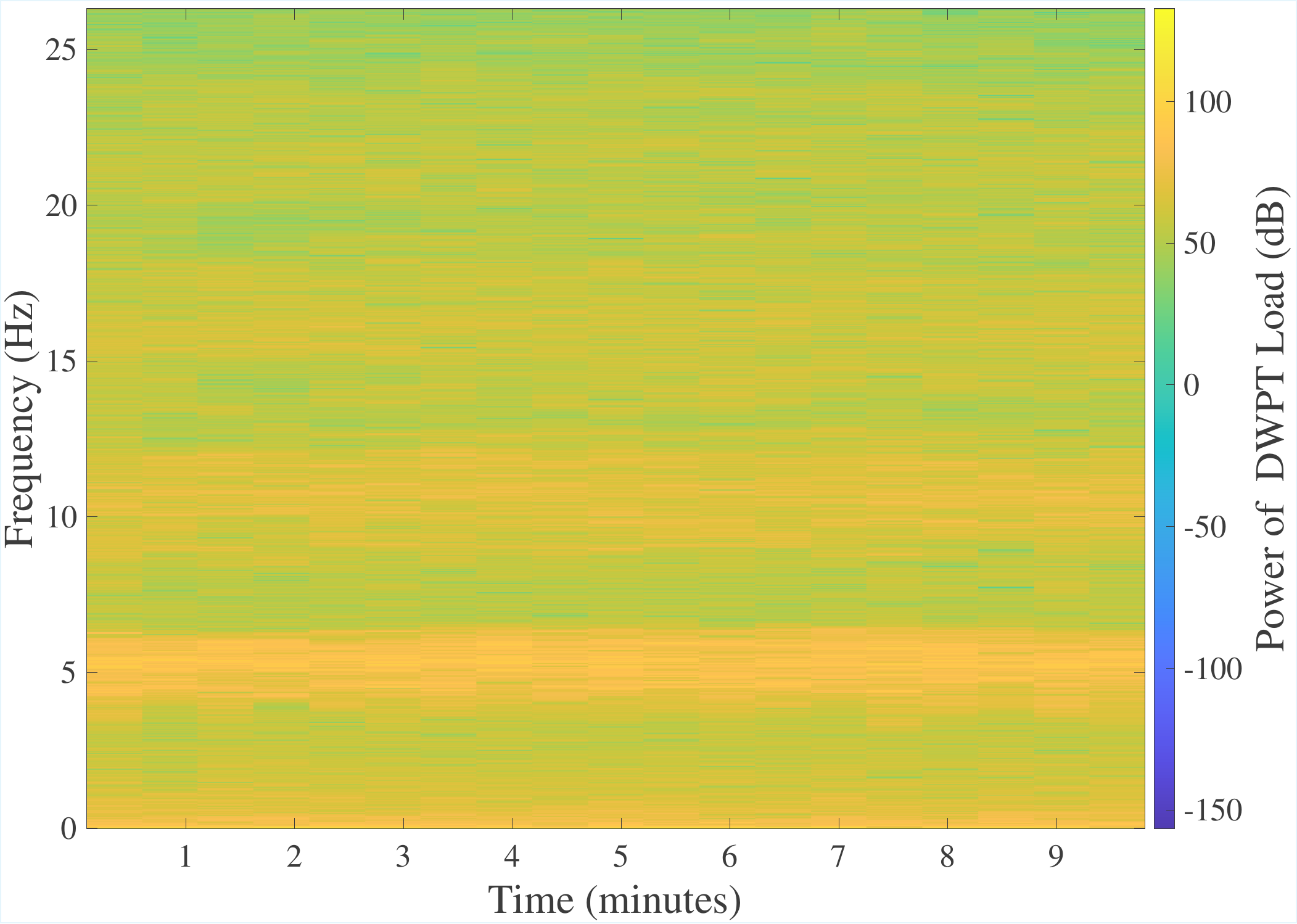}
\caption{Spectrogram of DWPT loads when EVs move in free-flowing conditions. Observe that frequency content remains approximately constant across time as all EVs maintain their speeds. The power in the signal is concentrated at around the harmonics as predicted under scenario \emph{S$_4$}. Moreover, we still observe some spread around harmonic peaks due to EVs moving at time-varying speeds.}
\label{fig:spectro_freeflow}
\end{figure}

We also computed the spectrogram of the DWPT load to study how its frequency content varies with time. As pointed out in Fig.~\ref{fig:spectro_freeflow} under free-flowing traffic conditions, any $1$-min realization of the DWPT load would yield a similar spectrum and THC, as all EVs maintain their speeds across time.

\emph{Congested traffic conditions.} With the next test, we wanted to see if there exist traffic conditions under which the DWPT demand exhibits higher THC due to EV synchronization as in \emph{S$_1$}. To this end, we studied the DWPT load on an ER with heavy traffic moving at a slower speed. Such conditions can arise from accidents, adverse weather, or construction activities. To simulate such traffic conditions in SUMO, the speed limit was $24.56$~m/s ($55$~mph) for the first half of the ER and reduced for the second half. Initial EV speeds were drawn from the truncated Gaussian distribution explained earlier. The mean arrival rate was set to $\lambda_a = 1$~EV/s. At this rate, the traffic jam propagated from the second half backward to the beginning of the ER. Eventually, all EVs were moving at roughly the slower speed limit of the second half. Inter-vehicle spacings were found to be approximately constant across vehicles and time. 

\begin{figure}[t]
\centering
\includegraphics[width=\linewidth]{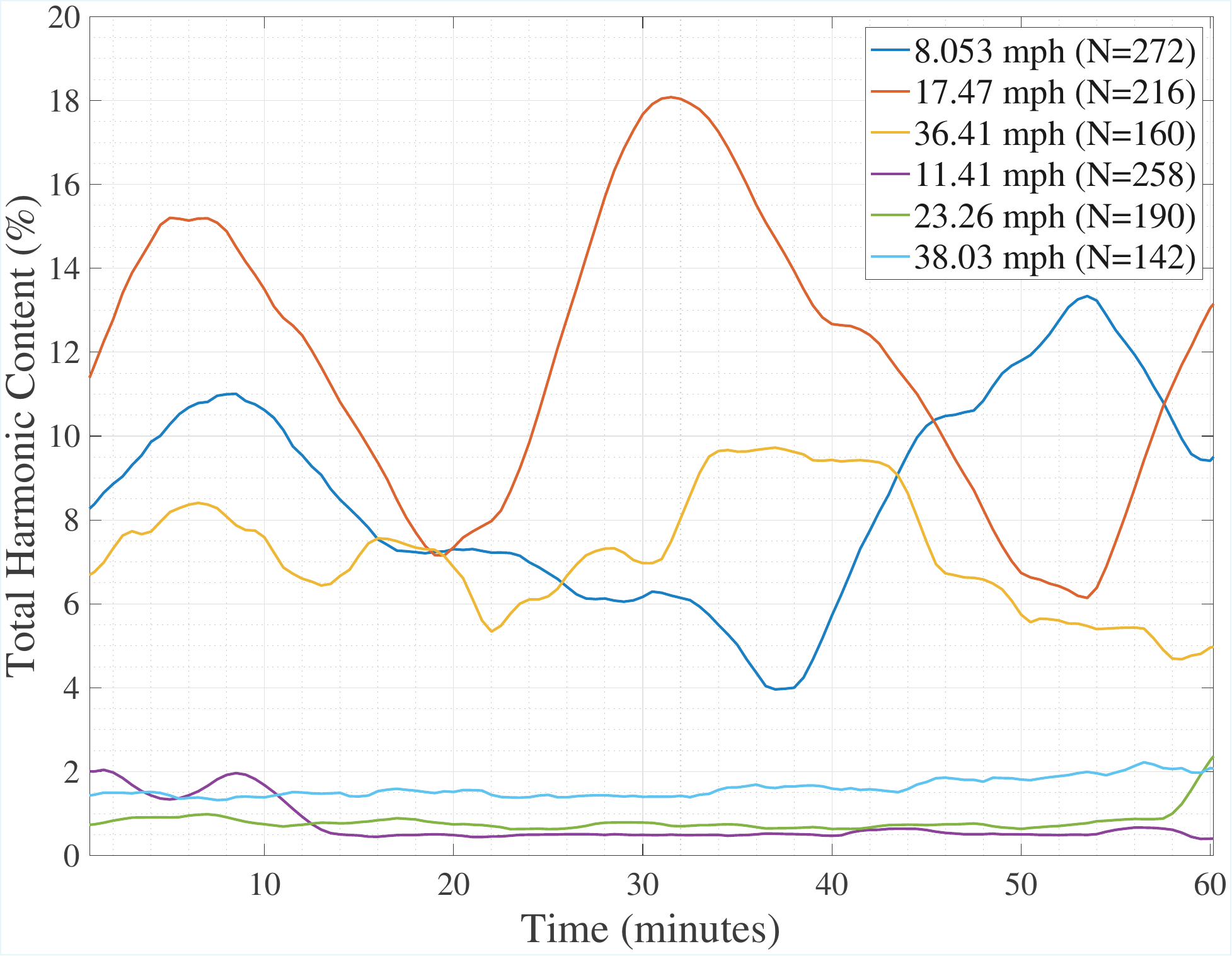}
\caption{Variation of THC across time under low-speed, high-volume traffic simulated in SUMO. Speed affects the number of EVs and their spacing. Some speeds result in near-synchronization and high THC, as in \emph{S$_1$}, and other speeds result in lower THC, as in \emph{S$_4$}.}
\label{fig:sumo_congestion}
\end{figure}

As explained under \emph{S$_1$}, EV loads at constant speed can synchronize only when EV spacings are integer multiples of the coil segment length $D$. Moreover, the distance gap between EVs increases with the speed limit~\cite{Krauss98}. Based on this understanding, we identified several speed limits for the second half of the ER, at which approximate synchronization occurs and leads to high THC. Analogous to Fig.~\ref{fig:spectro_freeflow},  Figure~\ref{fig:sumo_congestion} depicts how the THC of the total DWPT load demand evolves across time for different jam speeds. The THC was computed from $1$-min segments of $p(t)$ with 50\% overlap. We observe that higher speeds result in fewer EVs, as expected based on Remark~\ref{re:EVnumber}. We also note that the THC scales as $\text{THC}_h/\sqrt{N}$ for the bottom three speed limits, for which the inter-vehicle distance was not a multiple of $D$. On the contrary, for the top three speed limits, THC increases due to the resonance of traffic with the ER infrastructure. The maximum THC value we observed was $18\%$ for a particular $1$-minute segment when the jam moves at $7.81$~m/s ($17.47$~mph). {Such synchronization is impossible in free-flowing conditions, as EV spacings are randomized.} Overall, this experiment validates the conjecture made in \emph{S$_1$} that the DWPT load demand can exhibit high THC under heavy traffic conditions moving at specific speeds, especially at relatively lower speed limits. Also, the THC appears to be varying sinusoidally, which is a common phenomenon for frequency-modulated signals~\cite{proakis}. A detailed explanation of this sinusoidal variation in THC would be the topic of future study.

\section{Effects of DWPT Loads on a Power System} \label{sec:dynamics_simu}
We simulated the impact of DWPT loads on the WECC 179-bus system using the ANDES grid dynamic simulator~\cite{ANDES}. The WECC system has a base load of 60.78~GW and comprises 29 generators. To emulate the increasing penetration of inverter-based resources, the system's inertia was reduced by 40\%. The WECC system has unstable modes with negative damping ratios within the 0.1--0.7~Hz range~\cite{haozhu2025DC}. In summary, our tests on the WECC indicate that, under congested traffic conditions, DWPT loads can excite grid frequency dynamics when their fundamental frequency is close to a critical mode. Moreover, distributing DWPT loads across geographically distant buses can exacerbate frequency oscillations. DWPT loads during free-flowing conditions are unlikely to excite any dynamics. The tests are described in detail next.

\begin{table}[t]
\label{tab:load_wecc}
\centering

\caption{Frequency Content of DWPT Loads Powered by WECC}
{\fontsize{8pt}{11pt}\selectfont
\begin{tabular}{|l|l|l|}
\hline
\hline
  \textbf{Cases} & \textbf{Load THC (\%)}  & \textbf{Load fundamental frequency (Hz)}\\
     \hline
     \hline
     Case A & (3,~4,~5) & (5.38,~5.86,~4.40)\\
     \hline
     Case B & (15,~12,~12) & (1.50,~2.11,~2.56)\\
     \hline
     Case C & (3,~2,~3.5) & (1.75,~2.25,~3.1)\\
     \hline
     Case D & (14,~12,~12) & (0.69,~2.11,~2.56)\\
     \hline
     \hline
\end{tabular}}
\end{table}

In all tests, we considered DWPT systems connected at three buses. Although the bus locations varied per test as specified, the DWPT spectrum parameters were selected as shown in Table~\ref{tab:load_wecc}. Case A simulates DWPT loads under free-flowing traffic conditions with DC power being approximately 10\% of the nominal load demand of the hosting bus. Cases B-D simulate DWPT loads due to congested traffic with DC power being approximately 20\% of the nominal load demand of the hosting bus. In Case C, DWPT loads do not synchronize as EV spacings are not integer multiples of the slab length $D$. DWPT loads were generated using vehicle trajectories simulated by SUMO applied to the INDOT ER specifications. To ensure that the DWPT load does not alter the system's initial operating point or modes, we subtracted the DC term of the DWPT load from the original non-DWPT load at the three hosting buses. 

\begin{figure}[t]
\centering
\includegraphics[width=\linewidth]{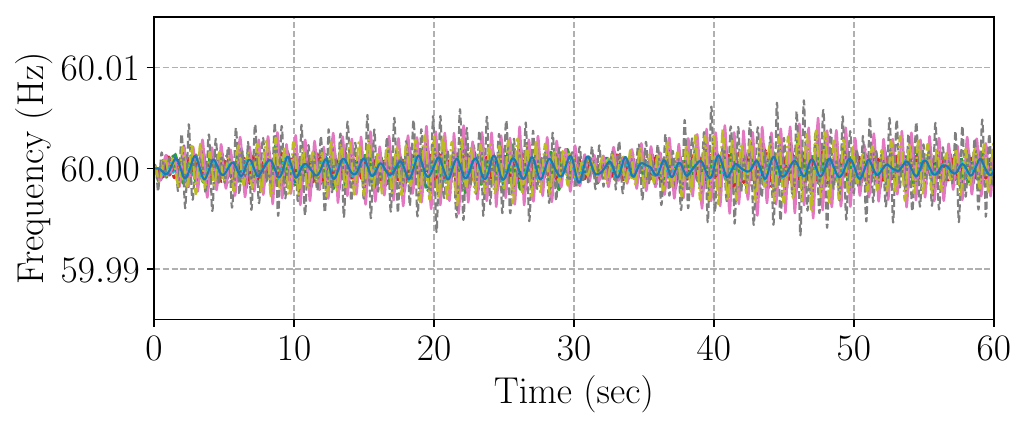}
\includegraphics[width=\linewidth]{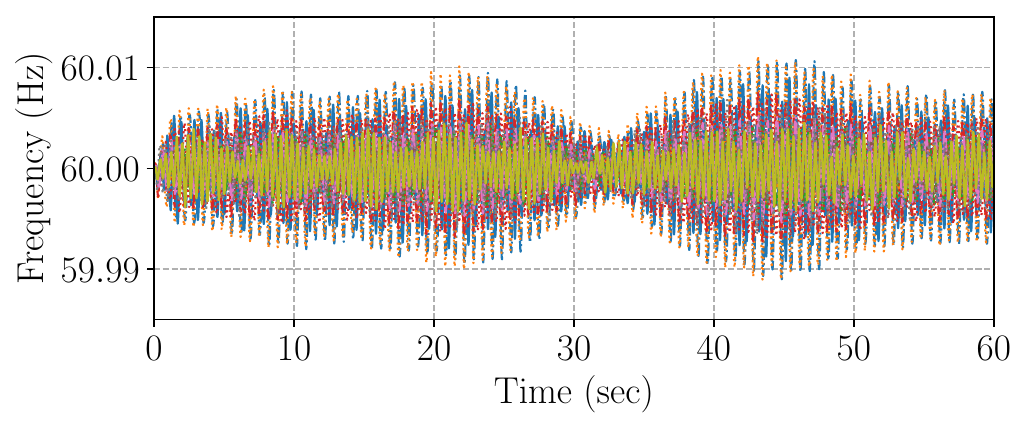}
\caption{Generator frequencies when DWPT loads are simulated per case B at: \emph{Top panel:} nearby buses (54,58,60); and \emph{Bottom panel:}) dispersed buses (54,100,164). The oscillations have a larger magnitude in the bottom panel; a maximum peak-to-peak deviation of 0.02 Hz, compared with 0.01 Hz in the top panel. One plausible explanation is that geographically distributed pulsating loads are more likely to excite multiple modes, thereby exacerbating oscillations.}
\label{fig:wecc_conc}
\end{figure}

\emph{Effect of DWPT location.} We first studied how the location of DWPT loads impacts grid dynamics by considering two sets of DWPT-serving buses: \emph{c1)} buses (54,58,60) that are geographically concentrated; and \emph{c2)} buses (54,100,164) that are geographically distributed. In either placement case, DWPT loads were simulated per Case B. Figure~\ref{fig:wecc_conc} shows generator frequencies in the respective cases. Although each DWPT load has the same DC power between the two cases,  case \emph{c2)} triggered stronger oscillations. In particular, we observed a maximum peak-to-peak frequency deviation of 0.01~Hz in \emph{c1)}, whereas in \emph{c2)}, peak-to-peak oscillations with an amplitude of 0.02~Hz are observed. A similar behavior of geographically dispersed pulsating loads yielding stronger oscillations has been observed in the context of pulsating datacenter loads while training large-language models (LLMs)~\cite{haozhu2025DC}.

\begin{figure}[t]
\includegraphics[width=\linewidth, keepaspectratio]{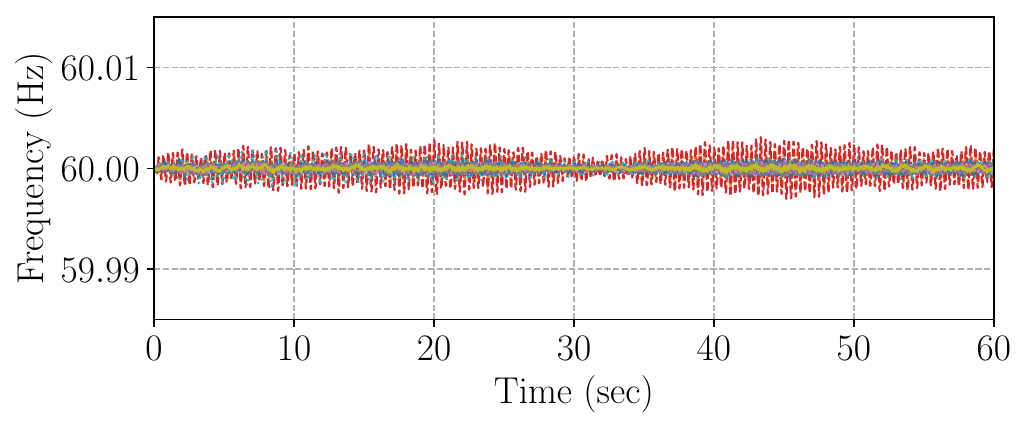}
\includegraphics[width=\linewidth, keepaspectratio]{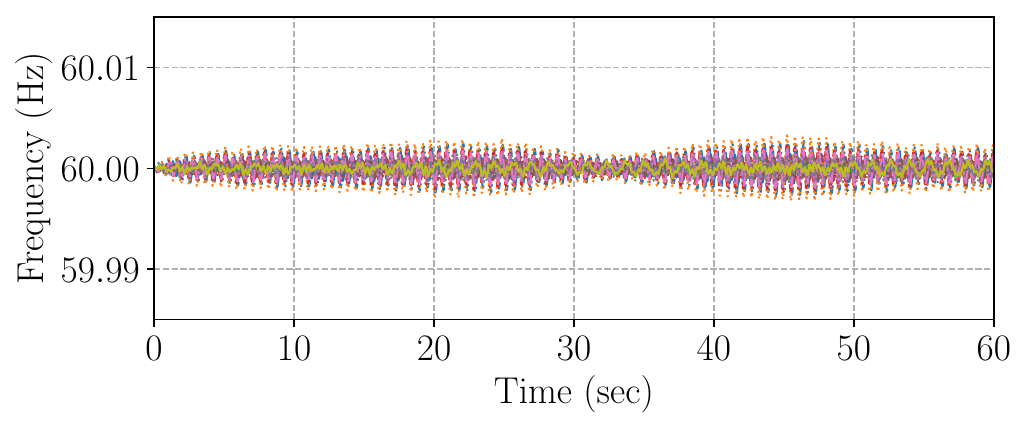}
\caption{Generator frequencies when DWPT loads are simulated per Case A (top) and Case C (bottom) at buses (54,100,164). Although loads are distributed geographically, their harmonic content is low, resulting in lower oscillation amplitudes.}
\label{fig:wecc_dist_caseAC}
\end{figure}

\emph{Effect of DWPT THC.} We simulated DWPT loads of different THCs sited on the same set of buses (54,100,164), and studied their impact on grid frequency dynamics. DWPT loads were simulated per Cases A, B, and C. Comparing the bottom panel of Fig.~\ref{fig:wecc_conc} with Fig.~\ref{fig:wecc_dist_caseAC}, we observe that DWPT loads exhibiting higher THC (Case B) introduce sustained frequency oscillations of amplitude 0.08~Hz. On the other hand, the lack of EV synchronization in Cases A and C introduces milder harmonics, leading to decaying oscillations.

\begin{figure}[t]
\includegraphics[width=\linewidth]{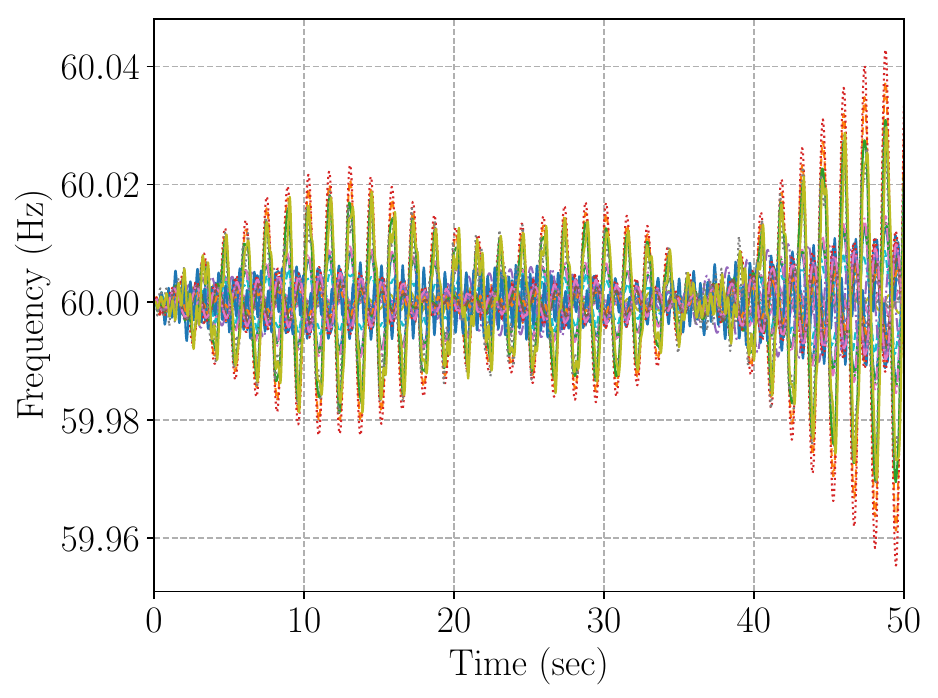}
\caption{Generator frequencies when DWPT loads are generated per case D at buses (54,100,164). Because one of the DWPT loads has a fundamental frequency close to the unstable 0.66-Hz mode, the system becomes unstable.}
\label{fig:wecc_dist_caseD}
\end{figure}

\begin{figure}[t]
\includegraphics[width=\linewidth]{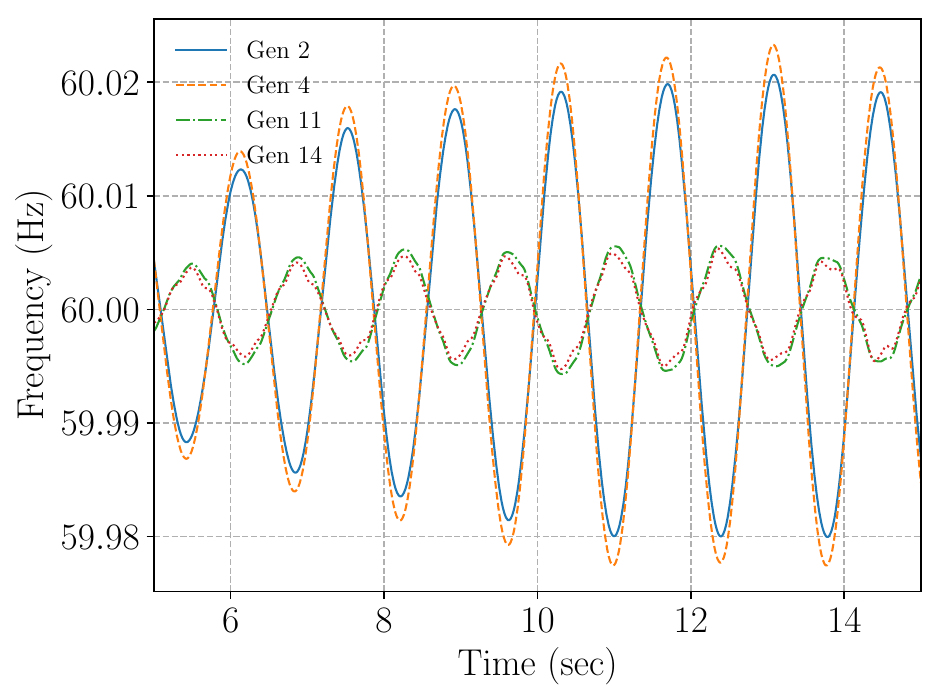}
\caption{DWPT loads generated per case D at buses (54,100,164) leading to inter-area oscillations. Heed generators (2,4) swing out of phase compared to generators (11,14). The two generator pairs are in different WECC zones.}
\label{fig:wecc_dist_interarea}
\end{figure}

\emph{Effect of DWPT fundamental frequency.} We also simulated grid dynamics when the DWPT load spectra contained peaks close to modal frequencies. The reduced-inertia WECC system exhibits a critical mode at 0.66 Hz, with significant participation from generator 14, located near bus 54. At traffic congestion speeds of around 8~mph, we observed that the DWPT load has a fundamental frequency of 0.69~Hz and a THC of 14\%; see the blue plot in Fig.~\ref{fig:sumo_congestion}. DWPT loads were generated per Case D at buses (54,100,160). The corresponding dynamics are shown in Fig.~\ref{fig:wecc_dist_caseD}. The DWPT load excites an unstable mode, causing the system to become unstable as generator frequencies diverge. Furthermore, frequencies associated with generators (2,4) swing out of phase compared to generators (11,14) as depicted in Fig.~\ref{fig:wecc_dist_interarea}. These sets of generators are in different WECC zones, indicating that the particular load configuration introduces inter-area oscillations.

\section{Conclusions}\label{sec:conclusion}
This work has developed statistical models for the total DWPT load to be consumed at the substation serving an ER. By postulating different distributions of EV timing, our analysis explains how the DWPT spectrum varies across different traffic conditions. The worst-case scenario for THC occurs when EVs move in a synchronized manner with specific inter-vehicle spacings. Simulations using SUMO demonstrate that near synchronization can occur under low-speed, high-volume traffic. When EVs move freely, the DWPT load exhibits a more favorable spectral profile, as confirmed by SUMO simulations of low-volume traffic. The frequency content at non-DC frequencies scales with $\sqrt{N}$ or $N$, depending on EV synchronism. Platoon formations can yield higher THC. In general, the fundamental frequency depends on the spacing between coils and the average EV speed, so low-speed traffic can result in harmonics below 2~Hz. Higher-order harmonics decrease in magnitude and spread across a wider bandwidth. Despite the simplicity of the models, our findings provide valuable insights to ER planners and grid operators on how the DWPT spectrum scales with the number of EVs. We also studied the impact of DWPT loads on the WECC system and identified that DWPT loads during congested traffic conditions could excite system dynamics. Geographic load spreading has been shown to worsen oscillations.

Some open research directions include: \emph{a)} Model DWPT loads as non-stationary signals and analyze their frequency content using tools such as the wavelet transform; \emph{b)} Study the effect of correlated EV speeds based on dynamical models of traffic flows; \emph{c)} Analyze the effect of DWPT load consumed at multiple buses on grid frequency dynamics; \emph{d)} Develop solutions to smooth out DWPT load, such as heterogeneous coil geometries; \emph{e)} Explore randomized triggering of Tx coils to naturally smooth out DWPT loads; and \emph{f)} develop solutions to actively monitor DWPT loads for efficient system operation.

\section{Appendix}\label{sec:appendix}
\begin{IEEEproof}[Proof of Lemma~\ref{le:s1}]
According to \eqref{eq:thc}--\eqref{eq:FS of p(t)}, the squared THC of $p(t)$ can be expressed as
\begin{equation}\label{eq:thcp2}
\text{THC}_p^2 = \frac{2\sum_{m=1}^{\infty}|\tilde{c}_m|^2}{\tilde{c}_0^2}=2\sum_{m=1}^{\infty}\frac{|\tilde{c}_m|^2}{\tilde{c}_0^2}.
\end{equation}
We would like to maximize THC$_p^2$ over the EV parameters $\{a_n,t_n\}_{n=1}^N$, collected in vectors $\ba$ and $\bt$, respectively. Define vector $\bb_m(\bt)\in\mathbb{C}^N$ so its $n$-th entry is $e^{-jm\omega_0t_n}$ for all $m\geq 0$. We can then express $|\tilde{c}_m|^2$ in \eqref{eq:thcp2} as
\[|\tilde{c}_m|^2=|c_m|^2\cdot |\ba^\top\bb_m(\bt)|^2.\]
Note that $\bb_0(\bt)=\bone$ for any $\bt$, where $\bone$ is the all-one vector. For any $m\geq 1$, consider maximizing the ratio over $(\ba,\bt)$: 
\begin{equation*}\label{eq:ratio}
\frac{|\tilde{c}_m|^2}{\tilde{c}_0^2}=\frac{|c_m|^2|\ba^\top\bb_m(\bt)|^2}{c_0^2|\ba^\top\bone|^2}
\leq 
\frac{|c_m|^2\|\ba\|_1^2\|\bb_m(\bt)\|_{\infty}^2}{c_0^2\|\ba\|_1^2}=\frac{|c_m|^2}{c_0^2}
\end{equation*}
due to H\"{o}lder's inequality and $\|\bb_m(\bt)\|_{\infty}=1$ for any $\bt$. 

Consider now some particular choices of $\bt$. Suppose all EVs have equal timings, so that $\bt=t_0\bone$ for some $t_0\in [0,T)$. For this $\bt$, it holds that $|\ba^\top\bb_m(\bt)|=|\ba^\top\bone|=\|\ba\|_1\cdot \|\bb(\bt)\|_{\infty}$, so the ratio $|\tilde{c}_m|/\tilde{c}_0$ attains its upper bound regardless of the value of $\ba$. Clearly, the choice of $\bt=t_o\bone$ maximizes $\text{THC}_p^2$ in \eqref{eq:thcp2}. The maximum value is $\text{THC}_p=\text{THC}_h$.
\end{IEEEproof}

\begin{IEEEproof}[Proof of Lemma~\ref{le:s2}]
First, apply the expectation operator on \eqref{eq:pnfs} to find the mean of $p_n(t)$ as
\begin{align}
\expect[p_n(t)] &= \expect\left[a_n \sum_{m=-\infty}^{\infty} c_m e^{jm\omega_0t}e^{-jm\omega_0t_n}\right]\nonumber\\
&=\expect[a_n]\sum_{m=-\infty}^{\infty}c_m e^{jm\omega_0 t}\cdot \expect[e^{-jm\omega_0t_n}]\nonumber\\
&=\mu_a\sum_{m=-\infty}^{\infty}c_m e^{jm\omega_0 t}\delta[m] =\mu_a c_0\label{eq:proofmean}
\end{align}
where $\delta[m]$ is Kronecker delta function. The third equality holds due to $t_n$ being uniform in $[0,T)$ with $T=\frac{2\pi}{\omega_0}$ so that:
\[\expect[e^{-jm\omega_0t_n}]=\frac{1}{T}\int_{0}^Te^{-jm\omega_0t_n}\mathrm{d} t_n=\delta[m].\]
The mean can be computed as $\expect[p(t)] = \sum_{n=1}^N \expect[p_n(t)] = N\mu_a c_0$. The autocorrelation function of $p(t)$ can be found as
\begin{align}
\expect[p(t)p(t+\tau)]&= \sum_{n=1}^N \sum_{\ell=1}^N \mathbb{E}[p_n(t)p_{\ell}(t+\tau)]\nonumber\\
&=R_a(\tau)+R_c(\tau)\label{eq:Rtau}
\end{align}
The double summation involves $N$ terms with $\ell=n$ contained in $R_a(\tau)$, and $N(N-1)$ terms with $\ell\neq n$ contained in $R_c(\tau)$. We expound on these two terms, starting with the first one:
\begin{align*}
R_a(\tau)&= \sum_{n=1}^N \sum_{\ell=n} \mathbb{E}[p_n(t)p_n(t+\tau)]\\
&=N\mathbb{E}[p_n(t)p_n(t+\tau)]=N R_n(\tau).
\end{align*}
The autocorrelation of $p_n(t)$ can be computed as
\begin{align*}
R_n(\tau)&=\expect[a_n^2]\sum_{m}\sum_{k}c_m c_k e^{jm\omega_0t}e^{jk\omega_0(t+\tau)}\expect[e^{-j(m+k)\omega_0t_n}]\\
&=(\mu_a^2+\sigma_a^2)\sum_{m}\sum_{k}c_mc_ke^{j(m+k)\omega_0 t}e^{jk\omega_0 \tau}\delta[m+k]\\
&=(\mu_a^2+\sigma_a^2)\sum_{m}c_mc_{-m}e^{-jm\omega_0 \tau}\\
&=(\mu_a^2+\sigma_a^2)\sum_{m=-\infty}^{\infty}c_m^2e^{-jm\omega_0 \tau}\\
&=(\mu_a^2+\sigma_a^2)\left(c_0^2 +2\sum_{m=1}^{\infty}c_m^2\cos(m\omega_0\tau)\right).
\end{align*}

Since the loads $p_n(t)$ are uncorrelated, the term $R_c(\tau)$ is
\begin{align*}
R_c(\tau)&= \sum_{n=1}^N \sum_{\ell\neq n} \mathbb{E}[p_n(t)p_{\ell}(t+\tau)]\\
&= \sum_{n=1}^N \sum_{\ell\neq n} \expect[p_n(t)]\cdot \expect[p_{\ell}(t+\tau)]\\
&=N(N-1) \mu_a^2 c_0^2,
\end{align*}
where the third equality follows from \eqref{eq:proofmean}. Adding $R_a(\tau)$ and $R_c(\tau)$, and collecting the terms related to the DC term, provides the final expression for the autocorrelation of $p(t)$:
\begin{equation*}
R_p(\tau)=  (N^2\mu_a^2 + N\sigma_a^2)c_0^2+ 2N(\mu_a^2+\sigma_a^2)\sum_{m=1}^{\infty} c_{m}^2 \cos(m\omega_0\tau).
\end{equation*}
The signal $p(t)$ is WSS because its mean is time-invariant and its autocorrelation depends only on $\tau$. Its PSD $S_p(\omega)$ can be found as the Fourier transform of $R_p(\tau)$.
\end{IEEEproof}

\begin{IEEEproof}[Proof of Lemma~\ref{le:s3}]
We index platoons by $i=\{1,\ldots,N/Q\}$. Let the EVs belonging to platoon $i$ constitute the set $\mcN_i$ and share timing $t_i$, which is drawn uniformly at random within $[0,T)$. The total load can be decomposed as
\begin{align*}
p(t)&=\sum_{i=1}^{N/Q} \sum_{n\in\mcN_i}p_n(t)=\sum_{i=1}^{N/Q}\left(\sum_{n\in\mcN_i}a_n\right)\sum_{m}c_me^{jm\omega_0(t-t_i)}\\
&=\sum_{i=1}^{N/Q}a_i\sum_{m}c_me^{jm\omega_0(t-t_i)}
\end{align*}
where $a_i$ is defined as the effective rating of platoon $i$. Based on this decomposition, the DWPT load is now amenable to the analysis of Lemma~\ref{le:s2} subject to two modifications. First, the signal $p(t)$ is now the sum of $N/Q$ rather than $N$ stochastic components, each with i.i.d. uniformly distributed timing $t_i$. Second, the effective rating $a_i$ of each stochastic component has a mean $Q\mu_a$, and its variance is increased to $Q\sigma_a^2$. The mean and autocorrelation of $p(t)$ can be obtained as in Lemma~\ref{le:s2} by simply replacing $N$ by $N/Q$, $\mu_a$ by $Q\mu_a$, and $\sigma_a^2$ by $Q\sigma_a^2$. The autocorrelation function is
\begin{align*}
R_p(\tau) &= (N^2\mu_a^2 +N\sigma_a^2)c_0^2\\ 
&\quad\quad+ 2\left(N^2\mu_a^2+ N\sigma_a^2\right)\sum_{m=1}^{\infty} c_m^2 \cos(m\omega_0\tau).
\end{align*}
The PSD can be found as the Fourier transform of $R_p(\tau)$.
\end{IEEEproof}

\begin{IEEEproof}[Proof of Lemma~\ref{le:s4}] This proof builds on the proof of Lemma~\ref{le:s2}. Applying the expectation operator on \eqref{eq:p(t)_unequalvel} yields
\begin{align*}
\expect [p(t)] &=  \sum_{n=1}^N \expect[a_n]\sum_{m=-\infty}^{\infty}  c_m\expect[e^{jm\omega_n(t-t_n)}] \\
&=  \sum_{n=1}^N \mu_a\sum_{m=-\infty}^{\infty}  c_m\delta[m] = N \mu_a c_0.
\end{align*}
The second equality stems from the law of total expectation, according to which~\cite{proakis}
\begin{align*}
    \expect[e^{jm\omega_n(t-t_n)}] &= \expect_{\omega_n}\left[\expect_{t_n}[e^{jm\omega_n(t-t_n)}| \omega_n]\right] \\ 
    &=\expect_{\omega_n} \left[ e^{jm\omega_nt} \expect_{t_n} [e^{-jm\omega_nt_n}| \omega_n]\right] \\
    &= \expect_{\omega_n} [e^{jm\omega_nt} \delta[m]] = \delta[m]
\end{align*}

Regarding the autocorrelation function, we follow the steps of Lemma~\ref{le:s2} and decompose $R_p(\tau)$ as [cf.~\eqref{eq:Rtau}]
\[R_p(\tau)=R_a(\tau)+R_c(\tau)=N R_n(\tau)+ N(N-1) \mu_a^2 c_0^2.\]
To find $R_p(\tau)$, it suffices to find the autocorrelation function $R_n(\tau)$ of $p_n(t)$ as follows:
\begin{align*}
&R_n(\tau)= \expect[a_n^2] \sum_{m,k}  c_m c_k \expect \left[e^{jm\omega_n (t-t_n)}e^{jk\omega_n (t-t_n)}e^{jk\omega_n\tau} \right]\\
&= \expect[a_n^2] \sum_{m,k} c_m c_k \expect_{\omega_n} \left[e^{j(m+k)\omega_nt} e^{jk\omega_n\tau}  \expect_{\tau_n} [e^{-j(m+k)\omega_nt_n}]\right] \\
    &= (\mu_a^2 + \sigma_a^2) \sum_{m,k} c_m c_k \expect_{\omega_n} [ e^{j(m+k)\omega_n t} e^{jk\omega_n \tau} \delta[m+k]] \\
    &= (\mu_a^2 + \sigma_a^2)  \sum_{m} c_m c_{-m} \expect_{\omega_n} [e^{-jm\omega_n \tau}]\\
    & = (\mu_a^2 + \sigma_a^2)  \sum_{m} c_m^2 e^{-jm\mu_{\omega}\tau} e^{-\frac{1}{2}m^2\sigma_{\omega}^2\tau^2}.
\end{align*}
The last equality follows from the characteristic function of the Gaussian distribution. In detail, for a Gaussian distributed random variable $x\sim\mcN(\mu_x,\sigma_x^2)$, its characteristic function can be shown to be~\cite{proakis}
\begin{equation*}
\Phi_{x}(z)=\expect[e^{jxz}]=e^{jz\mu_x}e^{-\frac{1}{2}z^2\sigma_x^2}.
\end{equation*}
Substituting $R_n(\tau)$ into $R_p(\tau)$ and separating the component related to $c_0^2$ provides the final expression for $R_p(\tau)$:
\begin{align*}
R_p(\tau) &= (N^2\mu_a^2 +N\sigma_a^2)c_0^2 \\ 
& \quad\quad +N(\mu_a^2+\sigma_a^2)\sum_{m\neq0} c_m^2 e^{-jm\mu_{\omega}\tau} e^{-\frac{1}{2}\sigma_{\omega}^2m^2\tau^2}.
\end{align*}
The PSD of $p(t)$ is the Fourier transform of $R_p(\tau)$. The Fourier transform of the constant term related to $m=0$ in $R_p(\tau)$ yields the $\delta(\omega)$ term in $S_p(\omega)$. Let us compute the Fourier transform of the terms related to $m\neq0$ in $R_p(\tau)$. To this end, we evaluate the next integral as
\begin{equation*}
\int_{-\infty}^{\infty} e^{-jm\mu_{\omega}\tau} e^{-\frac{1}{2}\sigma_{\omega}^2m^2\tau^2} e^{-j \omega \tau} d\tau=\frac{\sqrt{2\pi}}{m\sigma_{\omega}}e^{-\frac{1}{2}m^2\sigma_{\omega}^2\tau^2}.
\end{equation*}
This identity follows if we consider $\tau\sim \mcN(0,\frac{1}{m^2\sigma_{\omega}^2})$ and evaluate its characteristic function $\Phi_{\tau}(z)$ at $z=-(\omega+m\mu_{\omega})$. 
\end{IEEEproof}

\balance 

\bibliographystyle{IEEEtran}
\bibliography{myabrv,power,kekatos,DWPT}

\begin{IEEEbiography}[{\includegraphics[width=1in,height=1.25in,clip,keepaspectratio]{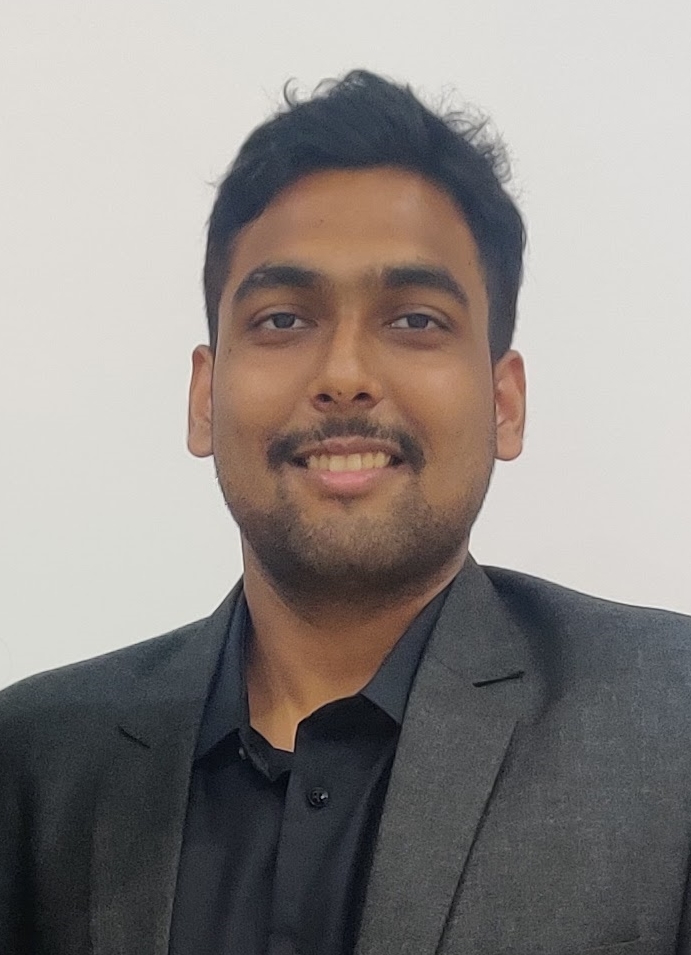}}] {Ashutossh Gupta} (GSM'24) received the B.Tech. degree from the Indian Institute of Technology (IIT) Bhubaneswar, India, in 2022; and the M.S. degree from Purdue University, West Lafayette, IN, USA, in 2026; both in Electrical Engineering. He is currently pursuing a Ph.D. degree at Purdue University. His research focuses on applying optimization, control, and signal processing techniques to analyze and control grid dynamics under the influence of large loads, such as data centers and electrified highways. 
\end{IEEEbiography}

\begin{IEEEbiography}[{\includegraphics[width=1in,height=1.25in,clip,keepaspectratio]{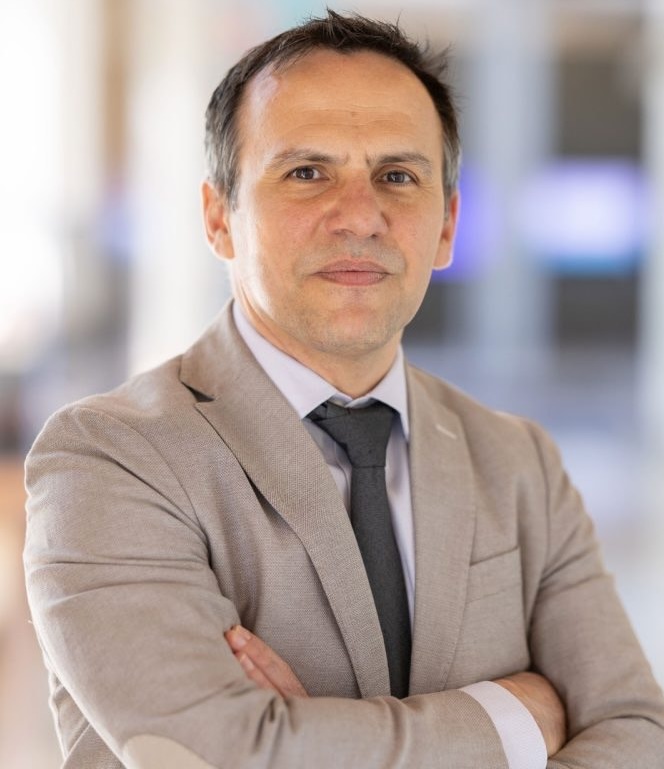}}] {Vassilis Kekatos} (SM'16) is an Associate Professor with the Schweitzer Power and Energy Systems group at the Elmore Family School of Electrical and Computer Engineering of Purdue University. He obtained his Ph.D. in Computer Science and Engineering from the Univ. of Patras, Greece, in 2007. He received a Marie Curie Fellowship from the European Commission during 2009-2012, and the US National Science Foundation CAREER Award in 2018. He was a postdoctoral research associate with the ECE Dept. at the Univ. of Minnesota. From 2015-2023, he was with the Bradley Department of ECE at Virginia Tech. From 2015 to 2022, he served as an Associate Editor for IEEE Trans. on Smart Grid, and now serves as an Associate Editor for IEEE Trans. on Power Systems. His current research focuses on optimization, machine learning, and quantum computing solutions for addressing power system computational tasks.
\end{IEEEbiography}

 \begin{IEEEbiography}[{\includegraphics[width=1in,height=1.25in,clip,keepaspectratio]{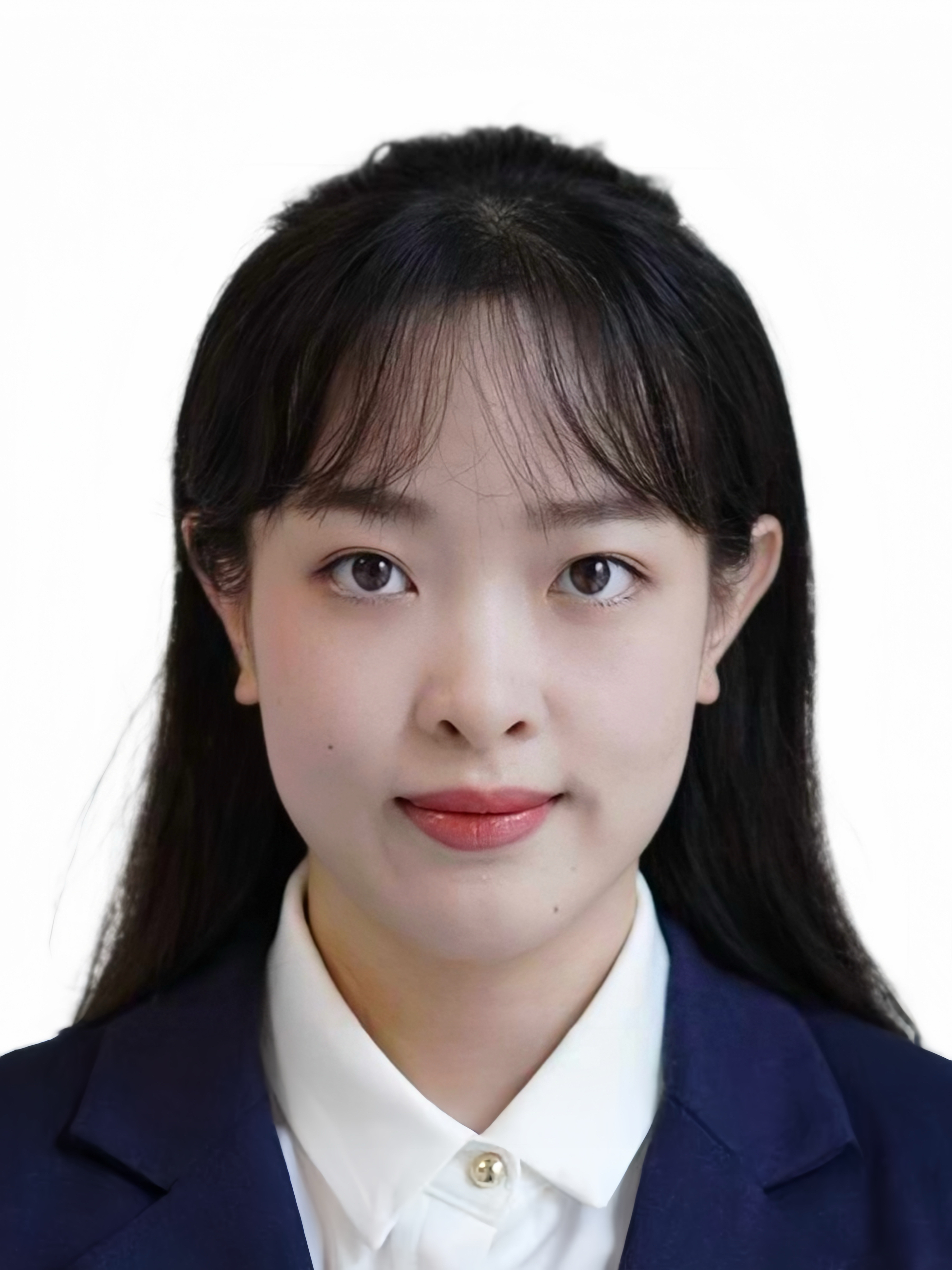}}]{Ruoyu Yang} received a B.E. degree in Electrical Engineering from Xi'an Jiaotong University, Xi'an, China, in 2023. She is now a Ph.D. student in the Elmore Family School of Electrical and Computer Engineering at Purdue University.
\end{IEEEbiography}

\begin{IEEEbiography}[{\includegraphics[width=1in,height=1.25in,clip,keepaspectratio]{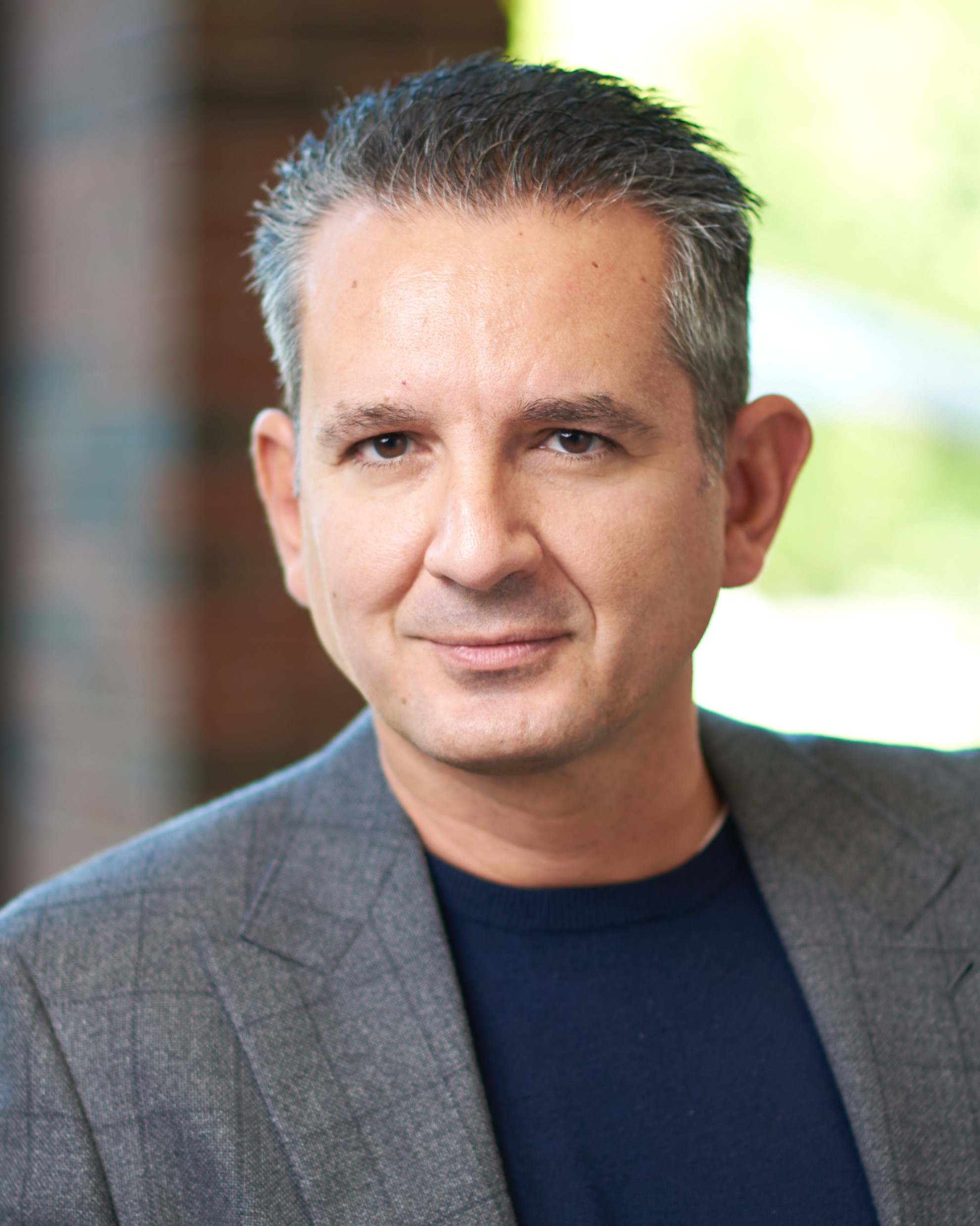}}] {Dionysios Aliprantis} (Fellow, IEEE) received the Diploma degree in electrical and computer engineering from the National Technical University of Athens, Athens, Greece, in 1999, and the Ph.D. degree from Purdue University, West Lafayette, IN, USA, in 2003. He is currently a Professor of Electrical and Computer Engineering at Purdue University. His research interests include electromechanical energy conversion and the analysis of power systems. More recently, his work has focused on technologies that enable the integration of renewable energy sources in the electric power system and the electrification of transportation. Prof. Aliprantis was the recipient of the NSF CAREER Award in 2009. He served as Editor-in-Chief for the IEEE Transactions on Energy Conversion.
\end{IEEEbiography}

\begin{IEEEbiography}[{\includegraphics[width=1in,height=1.25in,clip,keepaspectratio]{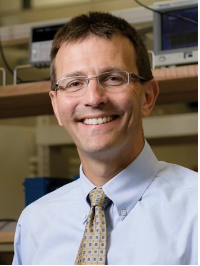}}] {Steve Pekarek} (Fellow, IEEE) received his PhD in Electrical Engineering from Purdue University in 1996. From 1997-2004, Dr. Pekarek was an Assistant (Associate) Professor of Electrical and Computer Engineering at the University of Missouri-Rolla. He is presently the Edmund O. Schweitzer III Professor of Electrical and Computer Engineering at Purdue University. He is an active member of the IEEE Power Engineering and Power Electronics Societies, the Electric Ship Research and Development Consortium (ESRDC), and the Research Director of the ASPIRE Center. He has served as the Program Chair of several IEEE conferences, including the International Electric Machines and Drives Conference and the Applied Power Electronics Conference. He is presently serving as the Vice President of Conferences for the IEEE Transportation Electrification Council.  
\end{IEEEbiography}
\end{document}